\documentclass{aa}
\usepackage{txfonts}
\usepackage{color}
\usepackage{graphicx}
\usepackage{mathptmx}
\usepackage{longtable}
\usepackage{natbib}

\def\degr{\hbox{$^\circ$}}

\def\farcs{\hbox{$.\!\!^{\prime\prime}$}}
\let\ni=\noindent

\usepackage{t1enc}
\usepackage{times}
\usepackage[utf8]{inputenc}
\usepackage{color, colortbl}
\usepackage{graphicx}
\usepackage{footnote}
\usepackage{pifont}
\let\ni=\noindent
\usepackage{lscape}
\usepackage{booktabs}
\definecolor{LightGray}{gray}{0.9}
\definecolor{LightGray1}{gray}{0.8}

\def\degr{\hbox{$^\circ$}}

\def\farcs{\hbox{$.\!\!^{\prime\prime}$}}

\let\ni=\noindent

\begin{document}

\title{Warsaw Catalogue of cometary orbits: 119 near-parabolic comets\thanks{All catalogue tables given in appendices (Tables~\ref{tab:orbital_quality108_list}, 
\ref{tab:orbit_osculating_27}-\ref{tab:NG-parameters}, \ref{tab:orbit_original_27}-\ref{tab:orbit_original_59},
 \ref{tab:orbit_future_27}-\ref{tab:orbit_future_59}) will be made available also in electronic form at the CDS via anonymous ftp to {cdsarc.u-strasbg.fr} (130.79.128.5) or via  {http://cdsweb.u-strasbg.fr/cgi-bin/qcat?J/A+A/}}
}

\author{Ma{\l}gorzata Kr\'olikowska}

\institute{Space Research Centre of the Polish Academy of Sciences,
           Bartycka 18A, 00-716 Warsaw, Poland \\
           \email{mkr@cbk.waw.pl}\label{inst1}           }

\authorrunning{Ma{\l}gorzata Kr\'olikowska}
\titlerunning{Warsaw Catalogue of cometary orbits}

\offprints{M. Kr\'olikowska, \email{mkr@cbk.waw.pl}}

%\date{Received 17 December 2013/Accepted 23 May 2014}

\abstract
{The dynamical evolution of near-parabolic comets strongly depends on the starting values of the orbital elements derived from the positional observations. In addition, when drawing conclusions about the origin of these objects, it is crucial to control the uncertainties of orbital elements at each stage of the dynamical evolution.}
{I apply a completely homogeneous approach  to determine the cometary orbits and their uncertainties. The resulting catalogue is
suitable for the investigation of the origin and future of near-parabolic comets.}
{First, osculating orbits were determined on the basis of positional data.
Second, the dynamical calculations were performed backwards and forwards  up to 250\,au from the
Sun to derive original and future barycentric orbits for each comet. In the present investigation of dynamical evolution, 
the numerical calculations for a given object start from the swarm of virtual comets constructed
using the previously determined osculating (nominal) orbit. In this way, the uncertainties of orbital elements were derived at the end of numerical calculations.} 
{Homogeneous sets of orbital elements for osculating, original and future orbits are given. The catalogue of 119 cometary orbits constitutes about 70 per cent of all the first class so-called Oort spike comets discovered during the period of 1801--2010 and about 90~per~cent of those discovered in 1951-2010, for which observations were completed at the end of 2013. Non-gravitational (NG) orbits are derived for 45~comets, including asymmetric NG~solution for six of them. Additionally, the new method for cometary orbit-quality assessment is applied for all these objects.}
{}


\maketitle

\section{Introduction}\label{sec:Introduction}

This catalogue presents for the first time the osculating, original and future orbital elements of near-parabolic comets that were determined by myself and in their majority were further investigated in collaboration with Piotr A. Dybczy\'nski from the Astronomical Observatory Institute at Pozna\'n. In a series of papers, i.e. \citet[hereafter Paper~1]{kroli-dyb:2010}, \citet[Paper~2]{dyb-kroli:2011}, \citet[Paper~3]{kroli-dyb:2012} and \citet[Paper~4]{kroli-dyb:2013}, we focused on the problem of the origin of near-parabolic comets with original inverse semi-major axes inside the so-called Oort spike. Investigating the dynamical orbital evolution  of these observed Oort-spike comets to the previous perihelion passage, we showed that only part of them make their first visit into the planetary zone, that is, we found that barely 50 per cent of these comets have a previous perihelion distance below 15\,au (Paper~2). 

The catalogue presents a homogeneous set of osculating orbits of just such comets, the nongravitational effects were determined for more than 40 per cent of them. 
The full sample of 119 comets forming this catalogue constitutes almost 70 per cent of all first class so-called Oort spike comets (condition of $1/a_{\rm ori} <150\times 10^{-6}$\,au$^{-1}$ was taken with a spare) discovered during the period of 1801--2010, $\sim$90~per~cent of those discovered in the years 1951--2010, ~for which observations were completed at the end of 2013. The completeness of all three subsamples of these comets is presented in more detail in section~\ref{sec:Warsaw_Catalogue}. In the next years the catalogue will be supplemented by comets discovered a long time ago, and will be updated with newly discovered objects after 2010. In particular, the complete sample of comets discovered during the years 1901-1950 is under consideration as an independent project of the New Catalogue of One-Apparition Comets, the first part of which, that includes 38 comets with $1/a_{\rm ori} \le 0.000130$\,au$^{- 1}$ according to \citet[hereafter MWC\,08]{MWC08:2008}, is completed. 

The paper is organized as follows: in the next section I~shortly describe the methods and model of motion applied to orbital determinations. In some cases the non-gravitational (hereafter NG) orbits were determinable, thus a brief description of the adopted model of NG~acceleration is given in subsection~\ref{sub:ng_acceleration}. Since the new method for evaluating the quality of osculating cometary orbits is used in the present catalogue, the accuracy of the cometary orbit is discussed in some detail in subsection~\ref{sec_orbit_accuracy}. The last part of section~\ref{sec:orbit_determination} %(\ref{sub:orbits_original_future}) 
describes how the original and future orbits and their uncertainties were determined. Section~\ref{sec:Warsaw_Catalogue} are divided into four parts that deal with tabular parts of the catalogue given in appendices. Thus, this section includes a general description of the observational material of the analysed comets (Part~I of the catalogue), as well as osculating orbital elements (Part~II), and original and future orbital elements (Parts~III and IV). The article ends with concluding remarks on the accuracy of orbital solutions given in the catalogue and future plans (Sect.~\ref{sec_conclusions}).

This publication is accompanied by an online catalogue ({\tt ssdp.cbk.waw.pl/LPCs}) also providing entries to full swarms of original and future virtual comets (hereafter VCs, see Sect.~\ref{sub:orbits_original_future}) that formed the basis for the detailed analysis of dynamical evolution presented in Papers~1--4. Thus, this catalogue also allows one to construct and analyse the observed distribution of {\it Oort spike} comets and to investigate the problem of cometary origin.

\section{Methods and assumptions}\label{sec:orbit_determination}

For each comet from the catalogue, I determined the osculating nominal orbit (GR -- pure gravitational or NG if possible; see next subsection) based on the astrometric data, selected and weighted according to the methods described in great detail in Paper~1. This allowed me to construct a homogeneous sample of cometary osculating orbits as well as homogeneous samples of original and future orbits.

The equations of a comet's motion were integrated numerically using the recurrent power series method \citep{sitarski:1989,sitarski:2002}, taking into account perturbations by all the planets (additionally, Pluto was taken into account to be consistent with DE405/WAW) and including the relativistic effects. All orbital calculations performed for this catalogue were based on the Warsaw numerical ephemeris DE405/WAW of the solar system \citep{sitarski:2002}, consistent with a high accuracy with the JPL ephemeris DE405.

\subsection{Nongravitational model of motion applied to orbit determination}\label{sub:ng_acceleration}

To determine the NG cometary orbit the standard formalism
proposed by \citet[hereafter MSY]{marsden-sek-ye:1973} was used where the
three orbital components of the NG~acceleration acting on a comet
are scaled with a function $g(r)$ symmetric relative to perihelion:

\begin{eqnarray}
F_{i}=A_{\rm i}\cdot & g(r) & ,\qquad A_{\rm i}={\rm ~const~~for}\quad{\rm i}=1,2,3,\label{eq:g_r}\\
 & g(r) & =\alpha\left(r/r_{0}\right)^{-2.15}\left[1+\left(r/r_{0}\right)^{5.093}\right]^{-4.614},\label{eq:ng_std}
\end{eqnarray}

\noindent where $F_{1},\, F_{2},\, F_{3}$ are the radial, transverse
and normal components of the NG~acceleration, respectively, and the
radial acceleration is defined outward along the Sun-comet line. 
The normalization constant $\alpha=0.1113$ gives $g(1$~AU$)=1$;
the scale distance $r_{0}=2.808$~AU. From orbital calculations,
the NG~parameters $A_{1},A_{2}$, and $A_{3}$ were derived together
with six orbital elements within a given time interval (numerical
details are described in \citealp{krolikowska:2006a}). The standard
NG~model assumes that water sublimates from the whole surface of
an isothermal cometary nucleus. The asymmetric model of NG~acceleration
is derived by using function $g(r(t-\tau))$ instead of g(r(t)). Thus, this
model introduces an additional NG~parameter $\tau$ -- the time displacement
of the maximum of the $g(r)$ relative to the moment of perihelion
passage. 

\vspace{0.1cm}

In the present catalogue, NG~solutions were determined for 45~near-parabolic comets, including asymmetric NG~solution for six of them 
(C/1959~Y1, C/1990~K1, C/1993~A1, C/2007~N3, C/2007~W1 and C/2008~A1).
Moreover, 66 per cent of comets with small perihelion distance have NG~orbits in the catalogue (33 of 50; see samples A1 and A2 in Sect.~\ref{sec:Catalogue_general} and Table~\ref{tab:orbital_quality108_list}).

\begin{figure}
\includegraphics[width=8.8cm]{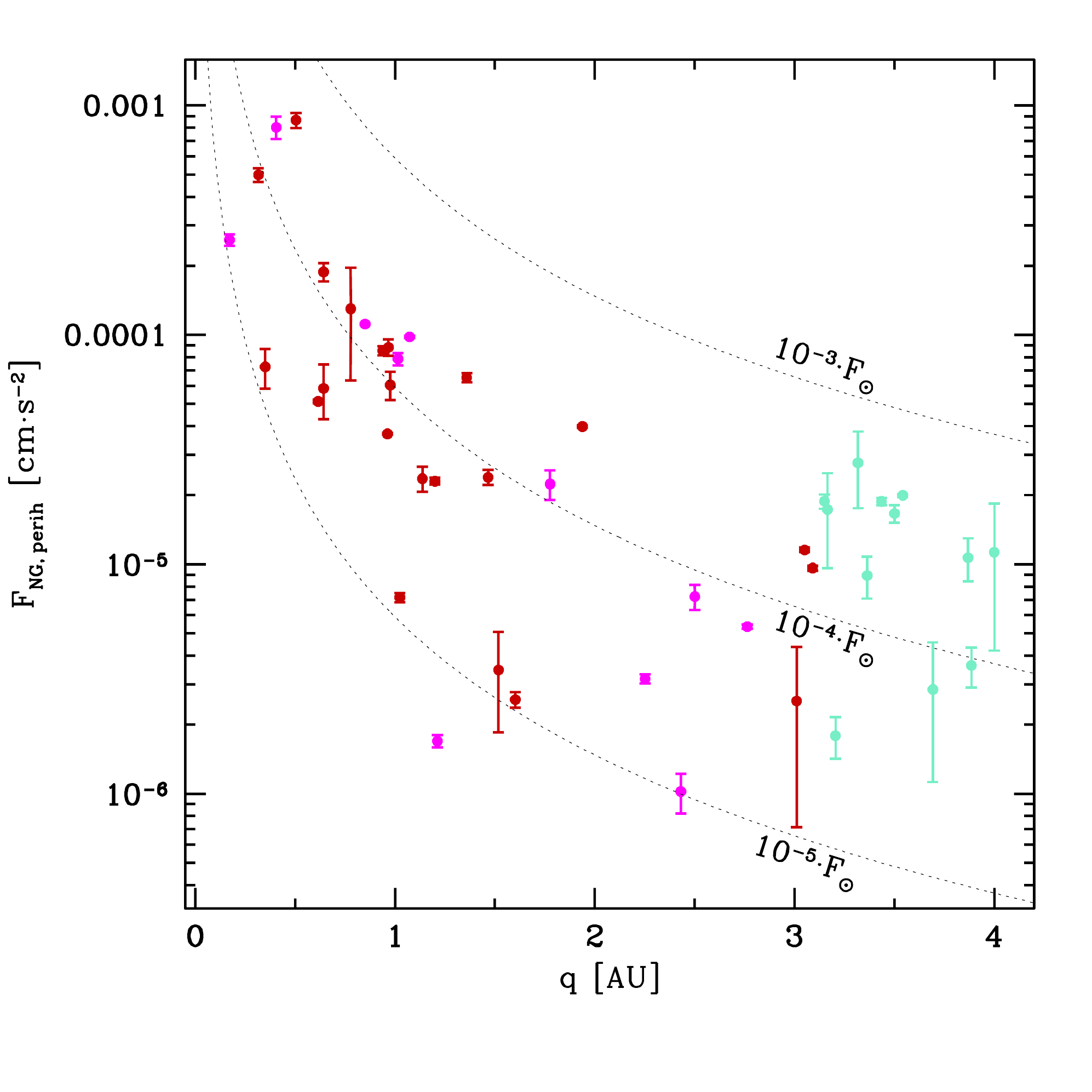} 
\caption{Maximum NG~accelerations, ${\rm F}_{\rm NG,perih}$, as a function of perihelion distance  
for all 45 near-parabolic comets with determinable NG~effects.
Three samples A1, A2 and B are shown by red, magenta and turquoise symbols, respectively.
The dotted curves represent the $10^{-5}$, $10^{-4}$ and $10^{-3}$ of the 
solar gravitational acceleration ${\rm F}_{\odot}$, respectively. NG~parameters taken for preparation of this plot 
are in Part~II of this catalogue (Table~\ref{tab:NG-parameters}).}
\label{fig:ng_force}
\end{figure}

NG~parameters for all NG~solutions given in this catalogue are presented in 
Table~\ref{tab:NG-parameters} in Part~II of this catalogue. 
One can see in Fig.~\ref{fig:ng_force} that NG~accelerations at perihelion derived for near-parabolic comets are below the $10^{-3}$ of solar gravitational acceleration with a typical value of about $10^{-4}\cdot {\rm F}_{\odot}$.

\subsection{Accuracy of the cometary orbit}\label{sec_orbit_accuracy}

In 1978 \citeauthor{mar-sek-eve:1978}~(hereafter MSE) formulated the
recipe to evaluate the accuracy of the osculating cometary orbits
obtained from the positional data. They proposed to measure this
accuracy by the quantity $Q$ defined as

\begin{eqnarray}\label{eq:orbit_accuracy_2}
 &  & Q=Q^*+\delta, {\rm ~~where~~}\nonumber \\
 &  & Q^*=0.5\cdot(L+M+N) {\rm ~~~~and~}\delta=\>1{\rm ~~or~~}0.5   \\
 &  &  {\rm ~~~~~~~~~~~~~~~~~~~~~~~~~~~~~~~~~~~to~make~}Q{\rm ~an~integer~number,~and} \nonumber 
\end{eqnarray}

\noindent $L$ ~~~~~denotes a small integer number that depends on
the mean error of the determination of the osculating 1/a,

\noindent $M$ is a small integer number that depends on the time
interval covered by the observations, and

\noindent $N$ is a small integer number that reflects the number of
planets whose perturbation were taken into account.

\vspace{0.1cm}

\noindent Values of $L$, $M$ and $N$ are obtained following the
scheme presented in the original Table~II given by MSE. The integer
$Q$-value calculated from Eq.~\ref{eq:orbit_accuracy_2} should next be
replaced with the orbit quality class as follows: a value of $Q=9,8$
means orbit of orbital class~1A, ~$Q=7$~of class~1B, ~$Q=6$~of class~2A, ~$Q=5$~of class~2B, and ~$Q<5$ means lower than second-class orbit.

\vspace{0.2cm}

In Paper~4 we discussed three reasons for which we found that some modifications of the above recipe of orbital accuracy estimation should be done. Briefly, these are the following:

\begin{enumerate}
  \item In the modern orbit determination all solar system planets
  are always taken into account, therefore we always have $N=3$.
  \item Current cometary positional observations are generally of
  significantly higher precision than 30 years ago. Moreover, modern LPCs are
  often observed much longer in time than the four years predicted by MSE in their scheme.
  Thus, the possibility of arbitrarily low values of a mean error
  of $1/a_{\rm osc}$ and an arbitrarily long time span of observations
  should be included to the original Table~II given by MSE.
  \item Almost all orbits of currently discovered LPCs should be
  classified as quality class~1A using the MSE quality scheme. Therefore, a better diversification
  between orbit accuracy classes is necessary. We realized this
  postulate by new $\delta$-definition and introducing three
  quality classes 1a+, 1a and 1b instead of the former 1A and 1B.
\end{enumerate}

\noindent The final form of a new orbital quality scheme was
constructed after an inspection of orbital uncertainties and data
intervals in the sample of 22~comets discovered in the years
2006--2010 (Paper~4), and samples of near-parabolic comets from
Papers~1--3. The new scheme proposed in Paper~4 is based  on a slightly modified
Eq.~\ref{eq:orbit_accuracy_2}:

\begin{eqnarray}\label{eq:orbit_accuracy_3}
 &  & Q=Q^*+\delta, {\rm ~~where~~}\nonumber \\
 &  & Q^*=0.5\cdot(L+M+3) {\rm ~~~~and~~~~}\delta=\>0{\rm ~~or~~}0.5, \\ 
 &  &  {\rm ~~~~~~~~~~~~~~~~~~~~~~~~~~~~~~~~~~~to~make~}Q{\rm ~an~integer~number.} \nonumber 
\end{eqnarray}

\vspace{0.1cm}

\noindent To distinguish the proposed quality system from
the MSE~system, in Paper~4 we used the lower-case letters 'a' and 'b' in quality class
descriptions instead of the original 'A' and 'B' in the following way:
$Q=9$~--~class~1a+, $Q=8$~--~class~1a, $Q=7$~--~class~1b,
$Q=6$~--~class~2a, $Q=5$~--~class~2b, $Q=4$~--~class~3a class, $Q=3$~--~class~3b, and $Q\le2$~--~class~4, where $Q$ is calculated according to
eq.~\ref{eq:orbit_accuracy_3}. The quality classes ~3a, ~3b~ and ~4~
were not defined by MSE, but we adopted here the idea published by
\citet{IAU_MPC_Web2} as 'a logical extension to the MSE~scheme'.

\vspace{0.1cm}

\noindent How to calculate the quantities $L$ and $M$ is described
in Table~\ref{tab:mse78_II}, which is a simpler form of the original
Table~II given by MSE. We only introduced in this table the
possibility of arbitrarily low values of a mean error of $1/a_{\rm
osc}$ and an arbitrarily long time span of observations, 
and, as mentioned before, we completely removed the redundant 
column describing the number of planets taken into account in the 
orbit determination process. 
Instead, we set N=3 in Eq.~\ref{eq:orbit_accuracy_3}. Thus, the
mean error of $1/a_{\rm osc}$ smaller than 1 unit (i.e.
$1\times10^{-6}$\,au$^{-1}$) now gives $L=7$ and a time span of data
longer than 48~months results in $M=8$.

\vspace{0.1cm}

This new orbit quality scheme separates the orbits of very good
quality in MSE~system, 1A, into three quality classes in the new
system, where the lowest of orbits in class~1A ($Q^*$=7) in the MSE system are
classified as 1b in the new scheme.

\begin{table}
\caption{\label{tab:mse78_II} Quantities for establishing
accuracy of orbit. This version of the table is taken from Paper~4.}

\centering{}%
\begin{tabular}{@{}rr@{~,~}lr@{~,~}l}
\hline \hline 
$L$ \& $M$  & \multicolumn{2}{c}{{Mean error of 1/a$_{{\rm
osc}}$}} & \multicolumn{2}{c}{{time span of
observations}}\tabularnewline
 & \multicolumn{2}{c}{{in units of $10^{-6}$\,au$^{-1}$}} & \multicolumn{2}{c}{{in months or days}}\tabularnewline
\hline {8}  & \multicolumn{2}{c}{{}} &
\multicolumn{2}{l}{{~~~~$\geq$ 48 months}}\tabularnewline {7}  &
\multicolumn{2}{l}{{~~~~~~~~$<$ 1}} & {$[$24}  & {48$[$}
\tabularnewline {6}  & {$[$1}  & {5$[$}  & {$[$12}  & {24$[$}
\tabularnewline {5}  & {$[$5}  & {20$[$}  & {$[$ 6}  & {12$[$}
\tabularnewline {4}  & {$[$20}  & {100$[$}  & {$[$ 3}  & {6$[$}
\tabularnewline {3}  & {$[$100}  & {500$[$}  & {$[$1.5}  & {3$[$}
\tabularnewline {2}  & {$[$500}  & {2500$[$}  & {$[$23} days  & {1.5
months$[$} \tabularnewline {1}  & {$[$2\,500}  & {12\,500$[$}  &
{$[$12}  & {23$[$ days} \tabularnewline {0}  &
\multicolumn{2}{c}{{$\geq$ 12\,500}} & {$[$7}  & {12$[$}
\tabularnewline {-1}  & \multicolumn{2}{c}{{}} & {$[$3}  & {7$[$}
\tabularnewline {-2}  & \multicolumn{2}{c}{{}} & {$[$1}  & {3$[$}
\tabularnewline \hline
\end{tabular}
\end{table}

\subsection{Original and future orbits}\label{sub:orbits_original_future}

To calculate the original and future orbital elements  as well as their uncertainties 
(taken at 250\,au from the Sun where planetary perturbations are negligible) 
the dynamical calculations for swarms of starting osculating orbits 
were performed for each catalogue comet. Each swarm was constructed according to 
the Monte Carlo method proposed by \citet{sitarski:1998}, 
where the entire swarm fulfil the Gaussian statistics of fitting to
positional data used for a given osculating orbit determination (examples are in Paper~1). 
Each swarm consists of 5\,001\,VCs including the nominal orbit; 
this number of orbital clones gives a sufficient sample
to obtain reliable statistics at 250\,au from the Sun.
Values of uncertainties of original/future orbital elements were 
derived by fitting the distribution of a given orbital element 
of an original/future swarm of VCs to Gaussian distribution. All distributions
of orbital elements including $1/a$-distributions of analysed comets were still perfectly
Gaussian at 250\,au from the Sun.  

\section{The catalogue}\label{sec:Warsaw_Catalogue}

\subsection{Description and structure of tabular materials}\label{sec:Catalogue_general}

To avoid very long tables with dozens of entries in each row it was convenient to divide the tabular material into four general parts.

\begin{description}
\item [{\bf Part I. Description of the observational material}] \hfill \\
  Table~\ref{tab:orbital_quality108_list} describes the sets of positional observations taken for each orbit determination and gives the new quality assessment of derived osculating orbits using the scheme proposed in Paper~4. The presentation of considered comets is separated into three samples.
\begin{description}
  \item[Sample~A1] \hfill \\
  This sample consists of 28 near-parabolic comets of $q_{\rm osc}<3.1$\,au and $1/a_{\rm ori}<150\times 10^{-6}$\,au$^{-1}$ discovered before the year 2006. Sample~A1 is very incomplete and includes 40 per cent of all discovered LPCs that fulfil the above conditions (see Table~\ref{tab:Comets108_threesamples}). Most comets from this sample (except for six objects, see Table~\ref{tab:orbital_quality108_list}) have NG orbits, and these solutions are characterized by a clear decrease in root-mean-square error (rms) compared with the rms for GR orbits, only C/1974~F1 displays a slight decrease of rms for NG~solution. Orbital solutions for most comets belonging to this sample were used for dynamical evolution discussed in Papers\,1--3.  However, three comets analysed in these papers (C/1913~Y1, C/1940~R2 and C/1946~U1) are not presented here because they are included in the independent project of the New Catalogue of One-Apparition Comets from the Years 1901--1950, Part~I of which is just completed. In the present catalogue the new NG~solutions are given for C/1990~K1, C/1993~A1,  C/2002~T7 (POST type of data), C/2003~K4, and four new comets are added  to the sample analysed previously (C/2001~B1, C/2002~Q5, C/2003~T3, C/2005~E2, Sect.~\ref{sub:ng_acceleration} and Table~\ref{tab:orbital_quality108_list}.
  \item[Sample~A2] \hfill \\
  This is a complete sample of 22 comets of $q_{\rm osc}<3.1$\,au and $1/a_{\rm ori}<150\times 10^{-6}$\,au$^{-1}$ discovered
  in the period 2006--2010 (Table~\ref{tab:Comets108_threesamples}). NG~solutions are determined for eleven of them. 
  These comets were examined in Paper~4. 
  \item[Sample~B] \hfill \\
  This is an almost complete sample of 69~comets of $q_{\rm osc}>3.1$\,au and $1/a_{\rm ori}<150\times 10^{-6}$\,au$^{-1}$ discovered in the period 1970--2010; the missing four comets (C/2008~S3, C/2009~F4, C/2010~S1, and C/2010~U3) were still observed at the end of 2013. Orbits of six comets of Sample~B were redetermined using longer data intervals than those used in Papers~1--2  (C/1997~A1, C/1999~F1, C/1999~N4, C/1001~J4, C/2003~S3, and C/2005~K1, see notes 3--8 to Table~\ref{tab:orbital_quality108_list}) and the sample was enriched with ten comets in comparison to previous analyses (C/2005~L3, C/2006~S3, C/2007~D1, C/2007~VO$_{53}$, C/2008~FK$_{75}$, C/2008~P1, C/2009~P2, C/2009~U5, C/2010~D3, and C/2010~R1). Twelve comets from this sample have NG~orbits in the present catalogue. 
\end{description} 
\item [{\bf Part II. Osculating orbital elements}]\hfill \\
  Tables~\ref{tab:orbit_osculating_27}, \ref{tab:orbit_osculating_22} and \ref{tab:orbit_osculating_59} show heliocentric osculating orbital elements for all objects structured in accordance with Table~\ref{tab:orbital_quality108_list}. Table~\ref{tab:NG-parameters} gives the NG~parameters for all comets with NG~orbits in the catalogue.
\item [{\bf Part III. Original orbital elements}]\hfill \\
  Tables~\ref{tab:orbit_original_27}, \ref{tab:orbit_original_22} and \ref{tab:orbit_original_59} include the original barycentric orbits in the same order. 
\item [{\bf Part IV. Future orbital elements}]\hfill \\
  Tables~\ref{tab:orbit_future_27}, \ref{tab:orbit_future_22} and \ref{tab:orbit_future_59} present the future barycentric orbits.
\end{description}

The entire catalogue, that is, all tables given in appendices will be made available also through the Strasbourg Astronomical Data Centre (CDS).

\begin{table*}
\caption{\label{tab:Comets108_threesamples} Completeness of the sample of comets in the present catalogue in comparison to MWC\,08 as a function of the time distribution of 
cometary discovery (five periods given in Col.[1]) 
and the perihelion distance ($q_{\rm osc}<3.1$\,au or $q_{\rm osc}\ge3.1$\,au, Cols.\,[2]--[3] and [4]--[5], respectively).
The number of comets are given for objects of first-quality class orbits in MWC\,08 (1A or 1B), except for the A2~sample of comets,
where 3 objects with orbital quality class 2 and 3 are included; this is indicated by '$+$ 3' in Cols.\,[3] and [7]. The upper limit 
for $1/a_{\rm ori}$ is taken at 0.000150\,au$^{-1}$. All comets with an NG~orbit in MWC\,08 (in this case the $1/a_{\rm ori}$ 
is not given there) were also analysed and some of them have $1/a_{\rm ori}<0.000150$\,au$^{-1}$. 
The sample A1+A2 is complete in the time interval of 2006--2010; during the same period,  Sample B is almost complete -- four comets are still observable, one of which, C/2010 U3 Boattini,  will pass perihelion in 2019!}
\centering{}%
{\setlength{\tabcolsep}{9pt} 
\begin{tabular}{ccccccc}
\hline \hline 
% after \\: \hline or \cline{col1-col2} \cline{col3-col4} ...
 & \multicolumn{6}{c}{ N u m b e r ~~~o f ~~~c o m e t s ~~~i n ~~~t h e ~~~d
i f f e r e n t ~~~s a m p l e s}\tabularnewline 
Period       & \multicolumn{2}{c}{$q_{\rm osc}<3.1$\,au} & \multicolumn{2}{c}{$q_{\rm osc}\ge3.1$\,au} & \multicolumn{2}{c}{A l l ~~~c o m e t s}\tabularnewline 
of discovery & MWC\,08     & Samples A1+A2   & MWC\,08  & Sample B   & MWC\,08  & Samples A1+A2+B    \tabularnewline 
{$[1]$}      & {$[2]$}     & {$[3]$}         & {$[4]$}  & {$[5]$}     & {$[6]$}  & {$[7]$}           \tabularnewline 
\hline 
before 1901  & 11          & 2               & --       & --          & 11  & 2  \tabularnewline 
 1901--1950  & 26          & 0               & 6        & --          & 32  & 0  \tabularnewline
 1951--2000  & 25          &16               & 42       & 34          & 67  & 50 \tabularnewline
 2001--2005  & 9           &10               & 20       & 20          & 29  & 30 \tabularnewline
 2006--2010  & 12          & 19 $+$ 3        & 6        & 15          & 18  & 34 $+$ 3 \tabularnewline 
\hline 
 All         & 83          & 47 $+$ 3        & 74       & 69          & 157  & 116 $+$ 3 \tabularnewline 
\hline
\end{tabular}}
\end{table*}

\subsection{Part~I of the catalogue. Characteristics of the observational material}
\label{sec_observational_material}

This part of the catalogue consists of one extensive table where the description 
of observational material is given for each comet including the new quality-class 
assessment proposed in Paper~4 and briefly described in Sect.~\ref{sec_orbit_accuracy}.

\vspace{0.1cm}

\noindent The full list of catalogue comets is shown in  Table~\ref{tab:orbital_quality108_list} and is available at {\tt ssdp.cbk.waw.pl/LPCs}.

\vspace{0.1cm}

Table~\ref{tab:orbital_quality108_list} includes 
the detailed description of the observational data sets used for
osculating orbit determination (Cols.\,$[4]$--$[8]$), type of the best
possible model that can be determined using these data (Col.\,$[9]$) and then
its quality assessment in the proposed new scheme (Col.\,$[12]$).
The number of residuals used for orbit determination and resulting rms 
are given in Col.\,$[13]$, and Col.\,$[14]$ informs where this solution was used for dynamical studies.

Since Table~\ref{tab:orbital_quality108_list} is also focused upon the best/preferred orbit for investigating the dynamical origin of near-parabolic comets, 
the more dedicated models for this purpose are given whenever it is necessary, that is,
when even the NG~model based on entire data set not fully
satisfies three criteria described in detail in Papers~1--4
(based on rms, O-C-diagram and O-C-distribution). 
This is indicated by subscript 'un' in Col.\,$[9]$ of
Table~\ref{tab:orbital_quality108_list} (some additional details are
given in the notes to this table).  Thus, for the cases marked 
NG$_{\rm un}$ in Col.\,$[9]$, the better models based on some
subsample of data are shown in the next rows for a given comet.

\noindent For an unsatisfactory solution based on an entire data set, it was found that the 
best method is to divide the data set into pre- and
post- perihelion subsets to determine the pre-perihelion and
post-perihelion osculating orbit for the purpose of past and future
dynamical evolution, respectively (as for example for C/2007~W1), or 
to construct a dedicated subset of data taken at large heliocentric distances from the Sun
(as for C/2001 Q4). 
Then the preferred models for the past/future
dynamical evolution are given in the second/third row of a given object in
almost all cases where orbital solution based on the entire data seems
to be unsatisfactory (except for two comets, C/1990~K1 and C/1993~A1, where a more
individual data treatment has not been taken yet; see also notes to Table~\ref{tab:orbital_quality108_list} 
for these comets). It is important to note that in these cases
the quality assessment may change as, for example, for comet C/2007~W1 or C/2008~A1. 

\noindent Therefore, the statistics of orbital classes also depends on types of osculating
orbits taken into account. Table~\ref{tab:statistics} shows this statistics for the sample
of 119~near-parabolic comets for two sets of osculating orbits: the
best osculating orbits determined on the basis of an entire interval of
data (second column labelled as 'standard') and the most preferred orbits
for the past dynamical investigation (third column labelled as 'past
evolution').

\begin{table}
\caption{\label{tab:statistics} Statistics of orbital quality classes for investigated comets; more details in the text} 
\centering 
\begin{tabular}{ccc}
  \hline \hline
  % after \\: \hline or \cline{col1-col2} \cline{col3-col4} ...
  orbital class~~~~ & standard~~~~ & past evolution \\
  \hline  
      1a+ & 37 & 34 \\  % poprawione ze wzgledu na nowe rozwiazanie dla C/2002 J4 i dodanie 14 komet (2013 12 10)
      1a  & 46 & 47 \\  % poprawione ze wzgledu na nowe rozwiazanie dla C/2002 J4 i dodanie 14 komet (2013 12 10)
      1b  & 25 & 26 \\
      2a  &  8 &  9 \\
      2b  &  2 &  2 \\
      3   &  1 &  1 \\
 \hline
\end{tabular}
\end{table}

\noindent The osculating orbits corresponding to all models described in Table~\ref{tab:orbital_quality108_list} as well as their original and future orbits are given in Parts~II, III and IV, respectively. 

\subsubsection{Application of the new scheme of orbital quality assessment}
\label{sec_orbit_accuracy_application}

The new quality estimates of the orbits of 119~LPCs are presented in Cols.\,$[10]$--$[12]$ of
Table~\ref{tab:orbital_quality108_list}. 

There is a higher diversification in orbital classes
between investigated comets than was found using the original MSE~recipe. This
can easily be checked by taking the $Q^*$-values given in Col.\,$[10]$ or
$[11]$ of Table~\ref{tab:orbital_quality108_list} and using
Eq.~\ref{eq:orbit_accuracy_3}.

In Papers~1--4, the orbit determination is based quite often on a larger data set (in
most cases currently available at \citealp{IAU_MPC_Web}) than was used
in MWC\,08 or at \citeauthor{IAU_MPC_Web} to obtain the orbital elements given there. 
Thus, our orbital quality assessment sometimes lists a better quality class
than that shown in MWC\,08, although our method of quality
assessment is more restrictive (see numerous notes appended to
Table~\ref{tab:orbital_quality108_list}).

\subsubsection{Sample A1. Near-parabolic comets of $q_{\rm osc}<3.1$\,au discovered before the year 2006}
\label{sub_27comets}

The modified method of orbital quality determination was applied
to pure GR as well as to NG~orbits of analysed small perihelion comets. All Oort
spike comets were chosen from MWC\,08 as objects with highest
quality orbits (classes 1A or 1B) or with an NG~orbit\footnote{The quality
orbit assessment of near-parabolic comets with NG~orbits is not
given in MWC\,08}. 
  
\noindent Table~\ref{tab:orbital_quality108_list} shows that 
of the 28~comets of $q_{\rm osc}<3.1$\,au almost all
have determinable NG~orbits; C/1992~J1, C/2001~B1, C/2001~K3, C/2002~Q5, C/2003~T3, and C/2005~E2 are six exceptions.
We found that six comets from this sample should be classified as second-quality orbit
according to new, more restrictive method (see Table~\ref{tab:orbital_quality27}). 
Uncertainties of $1/a_{\rm osc}$ are lower than 20 in units
of $10^{-6}$\,au$^{-1}$ for five of them (Col.\,$[3]$ of Table~\ref{tab:orbital_quality27}), 
which means that they are significantly poorer than for the remaining comets in this
sample. All five have NG~orbits, which is a natural consequence because
more parameters were used for the orbit determination (six orbital elements plus  NG~parameters).
For the sixth comet, C/2001~K3, only a pure GR~orbit was determinable and 
its second-class quality of osculating orbit is mainly due to 
a very short time interval.  

\noindent Additionally, from seven comets with an NG~orbit in MWC\,08 only three listed in
Table~\ref{tab:orbital_quality27} have second-quality orbits
according to the modified method (applying the original MSE~method
to our orbit determinations -- only two comets). Some comment is
needed for C/1952~W1. In Paper~1, we determined an osculating orbit
from 36~positional observations taken from the literature\footnote{At the
\citet{IAU_MPC_Web} only six observations are available}. This is in
fact the only comet, for which we had determined the orbit on the basis
of fewer measurements than in the MWC\,08; for more details see
Paper~1. Thus, the MWC\,08 describes an orbit of C/1952~W1 as 1B
class (64 observations, the data interval is the same as ours),
whereas using the MSE~method for our orbit determination we obtain 
class~2A (Cols.\,2 and 3 of Table~\ref{tab:orbital_quality27}).

According to the more restrictive orbital quality
assessment in the sample of 28 comets we obtained 8 comets of class~1a+,
4~comets of class~1a, 10~of~class~1b, 5 of class~2a, and 1~object
of class~2b in this sample.

\subsubsection{Sample~A2. Near-parabolic comets of $q_{\rm osc}<3.1$ discovered in the years 2006--2010}
\label{sub_22comets}

In Paper~4 we selected all near-parabolic comets discovered in the
years 2006--2010 that have small perihelion distances, $q_{\rm
osc}<3.1$\,au, and $1/a_{\rm ori} < 0.000150$\,au$^{-1}$. According
to the proposed method of orbital quality assessment we have in this
sample a lower fraction of comets of the best orbital classes: 11
instead of 15 derived using the MSE quality system. Additionally,
these 11~comets are now divided into five~comets of class~1a+ and
six~comets of class~1a using the entire data sets shown in
Table~\ref{tab:orbital_quality108_list} in the first row of a given
object. Moreover, the orbit of one comet, C/2008~C1, is reclassified
as a second-class orbit. For more details see Paper~4.

\subsubsection{Sample~B. Near-parabolic comets of $q_{\rm osc}\ge3.1$\,au discovered in the period 1970--2010}
\label{sub_59comets}

In the sample of comets with large perihelion distances, 
the lowest value of $Q^{*}$ (eq.~\ref{eq:orbit_accuracy_3}) is 6.0 for two comets,
C/2006~YC and C/2007~Y1 (Table~\ref{tab:orbital_quality59}). 
Therefore, orbits of these two comets should be classified as second quality (2a class), according to our more restrictive method. In MWC\,08, both orbits 
were classified as class~2A probably because of the very short
intervals of observations, two months for C/2006~YC and 2.5 months for
C/2007~Y1 (Col.\,2 of Table~\ref{tab:orbital_quality59}). The current
orbits of these two comets are based on almost periods of data twice as long, 4 and 4.5 months, respectively, which resulted in improving the quality of orbits from second to 
first class when using the MSE method of quality assessment 
(Col.\,3 of Table~\ref{tab:orbital_quality59}). For this reason, these two
comets were also investigated by us in Paper~2. 

\noindent Three comets with greatest $1/a_{\rm osc}$-uncertainties in this sample,
C/1978~G2 ($37.9\times10^{-6}$\,au$^{-1}$), C/1983~O1
($20.5\times10^{-6}$\,au$^{-1}$) and C/1976~U1
($21.5\times10^{-6}$\,au$^{-1}$), now have class~1b, 1a, 1b (1B,
1A and 1B in MWC\,08), respectively, because of the relatively long time
interval of data. C/1983~O1 stands out among them because of the
longest data interval (7.5 years) and the closest perihelion distance
of 3.3\,au to the Sun, which allowed determining the NG~orbit (Paper~2). The
remaining two comets have $q_{\rm osc}>5.5$\,au, so any NG~traces in the motion, 
and data intervals shorter than two years. 
In general, the NG~orbit (if determinable) seems to be closer
to the actual motion of a given comet than the pure GR~orbit, but
when the NG~parameters are included in the orbit determinations this can
result in significantly grater uncertainties of orbital elements
than for the pure GR~orbit. This is the case of C/1983~O1, where
the uncertainty of $1/a$-determination of NG~orbit is one order of
magnitude larger ($20.5\times10^{-6}$\,au$^{-1}$) than for
GR~orbit ($1.5\times10^{-6}$\,au$^{-1}$). When NG~orbits, although
they are more realistic than pure GR~orbits, are characterized by 
significantly larger uncertainties of orbital elements than
GR~orbits, then the second parameter of orbital
quality assessment, the time interval of data, plays an important role,
as for C/1983~O1.

\begin{table*}
\caption{\label{tab:orbital_quality27} Comets of class~1A or 1B
or NG~orbits according to MWC\,08 and with orbital quality poorer
than first class according to the proposed, modified
classification of orbit quality for
28~small-perihelion comets ($q_{\rm osc}<3.1$\,au). Orbital class is not specified in MWC\,08 for comets
with NG~orbit given there; in Col.\,$[6]$ the number in parenthesis
shows a shorter time interval of data available when completing MWC\,08.}
\vspace{0.05cm}
\centering{}{\setlength{\tabcolsep}{4.7pt} %
\begin{tabular}{ccccccc}
\hline \hline 
Comet  & \multicolumn{3}{c}{Q u a l i t y ~~~o f ~~~o r b i t} & $1/a_{\rm osc}$-uncertainty  & data & References \tabularnewline
 & \multicolumn{1}{c}{MSE method} & \multicolumn{1}{c}{MSE method} & Modified method  & in units of  & interval & \tabularnewline
 & MWC\,08  & \multicolumn{2}{c}{a p p l i e d ~~~~t o ~~~~o u r ~~~~o r b i t s } & $[$ $10^{-6}$\,au$^{-1}$ $]$  & $[$ months $]$ & \tabularnewline
 & class  & Q (eq.~\ref{eq:orbit_accuracy_2}), class  & $Q^{*},\>Q($~eq.~\ref{eq:orbit_accuracy_3}), class  & our analysis  &  & \tabularnewline
$[1]$      & $[2]$     & $[3]$  & $[4]$         & $[5]$ & $[6]$  & $[7]$   \tabularnewline
\hline 
C/1885 X1  & NG orbit  & 7, 1B  & 6.0, 6.0, 2a  & 36.1  &  4.7   & Paper 1 \tabularnewline
C/1892 Q1  & 1B        & 7, 1B  & 6.0, 6.0, 2a  & 26.2  & 10.4   & Paper 1 \tabularnewline
%C/1940 R2  & 1B        & 7, 1B  & 6.0, 6.0, 2a  & 29.1  &  9.0   & Paper 1 \tabularnewline
C/1952 W1  & 1B        & 6, 2A  & 5.5, 6.0, 2a  & 187.5 &  7.3   & Paper 1 \tabularnewline
C/1959 Y1  & NG orbit  & 6, 2A  & 5.0, 5.0, 2b  & 101.5 &  5.5   & Paper 1 \tabularnewline
C/1989 Q1  & NG orbit  & 6, 2A  & 5.5, 6.0, 2a  & 25.9  &  4.1   & Paper 1 \tabularnewline
C/2001 K3  & 1B        & 7, 1B  & 6.0, 6.0, 2a  & 6.8   & 4.8 (3.0) & Paper 2 \tabularnewline
\hline
\end{tabular}}
\end{table*}

\begin{table*}
\caption{\label{tab:orbital_quality59} Same as in Table~\ref{tab:orbital_quality27} for large-perihelion comets ($q_{\rm osc}\ge3.1$\,au).}
\vspace{0.05cm}
\centering{}{\setlength{\tabcolsep}{4.7pt} %
\begin{tabular}{ccccccc}
\hline \hline 
Comet  & \multicolumn{3}{c}{Q u a l i t y ~~~o f ~~~o r b i t} & $1/a_{\rm osc}$-uncertainty  & data & References \tabularnewline
 & \multicolumn{1}{c}{MSE method} & \multicolumn{1}{c}{MSE method} & Modified method  & in units of  & interval & \tabularnewline
 & MWC\,08  & \multicolumn{2}{c}{a p p l i e d ~~~~t o ~~~~o u r ~~~~o r b i t s } & $[$ $10^{-6}$\,au$^{-1}$ $]$  & $[$ months $]$ & \tabularnewline
 & class  & Q (eq.~\ref{eq:orbit_accuracy_2}), class  & $Q^{*},\>Q($~eq.~\ref{eq:orbit_accuracy_3}), class  & our analysis  &  & \tabularnewline
$[1]$      & $[2]$     & $[3]$  & $[4]$         & $[5]$ & $[6]$  & $[7]$   \tabularnewline
\hline 
C/2006 YC  & 2A        & 7, 1B  & 6.0, 6.0, 2a  & 12.3  & 4.0 (2.0) & Paper 2 \tabularnewline
C/2007 Y1  & 2A        & 7, 1B  & 6.0, 6.0, 2a  & 12.4  & 4.6 (2.5) & Paper 2 \tabularnewline
\hline
\end{tabular}}
\end{table*}

\vspace{0.1cm}

According to our new orbital quality assessment, we obtained 24 comets of class~1a+,
36~comets of class~1a, 7~of~class~1b, and 2 objects of class~2a in sample~B.

\subsection{Part~II of the catalogue. Osculating orbital elements}\label{sec_osculating_orbits108}

This part of the catalogue consists of three tables (Tables~\ref{tab:orbit_osculating_27}, \ref{tab:orbit_osculating_22} and \ref{tab:orbit_osculating_59})
including samples A1, A2 and B, respectively. Each of them contains the comet designation and its name  (Col.\,$[1]$), epoch (osculation date, Col.\,$[3]$) and the six heliocentric orbital elements usually given for comets (Cols.\,$[3]$--$[9]$), that is, perihelion time (TT), perihelion distance (in astronomical units), eccentricity, argument of perihelion (in degrees, equinox 2000.0), longitude of ascending node (in degrees, equinox 2000.0), and inclination (in degrees, equinox 2000.0). Additionally, the inverse semi-major axis is presented in Col.\,$[9]$.

\vspace{0.1cm}

\noindent The astrometric data sets used for each orbital solution determination are presented in Part~I of the catalogue (Table~\ref{tab:orbital_quality108_list}).

\vspace{0.1cm}

\noindent The NG~parameters for all NG~osculating orbits included in the catalogue (i.e. for 45 comets, see Table~\ref{tab:orbital_quality108_list}) are shown in the separate Table~\ref{tab:NG-parameters}.
Most comets from sample~A1 have NG~orbits (almost 80~per cent). 
These NG~solutions are characterized by a clear decrease in rms compared with the rms for purely GR~orbits, 
only for C/1974~F1 this decrease is infinitesimal. Sample~A2 includes 11 osculating NG~orbits, 
thus 50 per cent of comets discovered in the period of 2006-2010
with a small perihelion distance ($q_{\rm osc}<3.1$\,au) have determinable NG~effects in their motion. 
On the other hand, only 12 of 69~comets of sample~B have NG~orbits.
For some comets even NG~orbits exhibit some trends in O-C~time variations. 
In these peculiar cases the second and sometimes the third orbital solutions 
are presented. Additional orbits are based on some subsamples of data, 
as described in Table~\ref{tab:orbital_quality108_list} (see also Sect.~\ref{sec_observational_material}). 
These dedicated solutions are more appropriate for studying the origin and/or future evolution of the cometary orbit.

\subsection{Part~III of the catalogue. Original orbital elements}\label{sec_original_orbits108}

Similarly as in Part~II, three tables (Tables~\ref{tab:orbit_original_27}, \ref{tab:orbit_original_22}, and \ref{tab:orbit_original_59}) corresponding to samples A1, A2, and B are presented here. 
Each of them contains the original barycentric orbits calculated at the distance of 250\,au from the Sun. 
Tables have an analogous structure and entries to those in Part~II. 

\begin{figure}
\includegraphics[width=8.8cm]{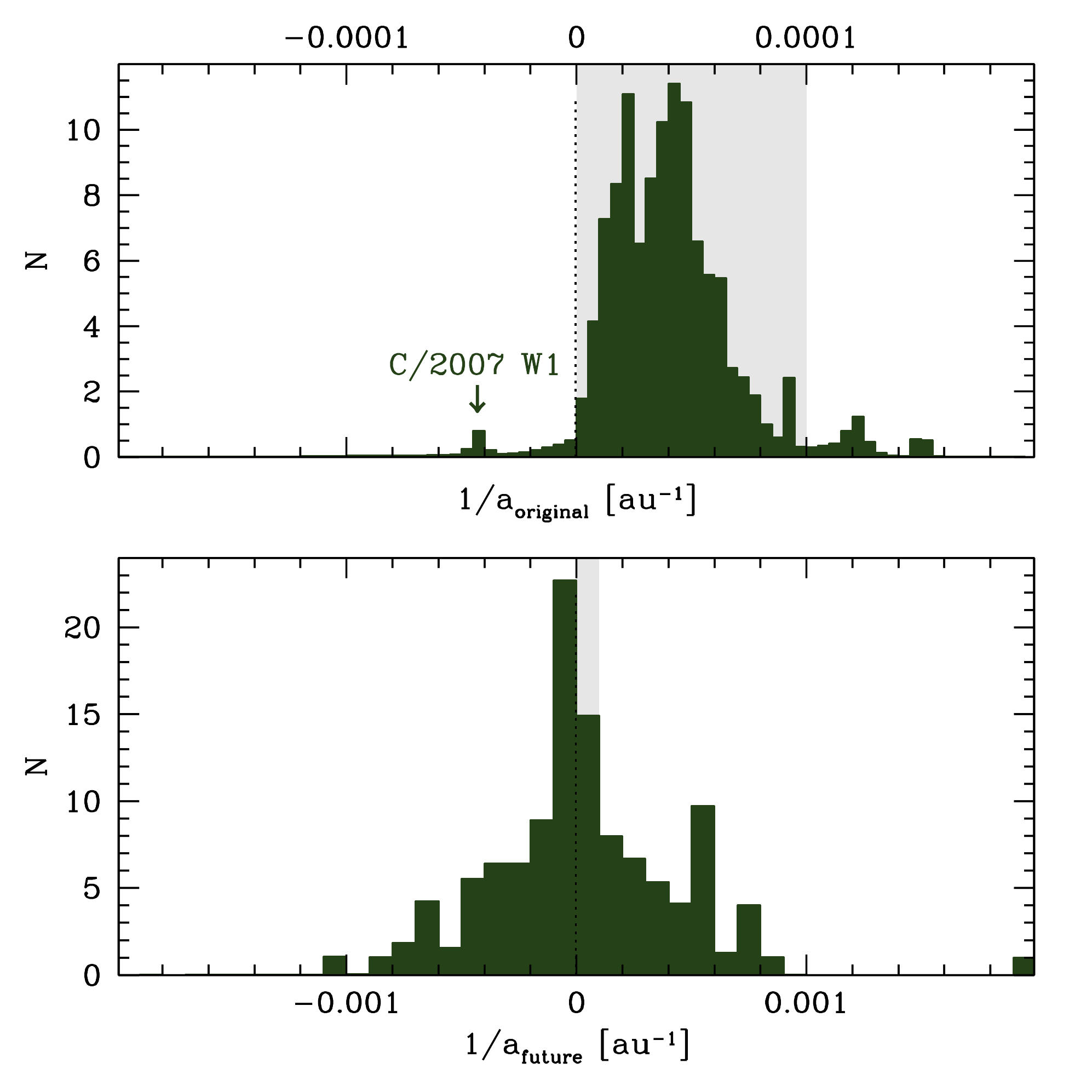}
\caption{Distribution of $1/a_{\rm ori}$ (upper panel), $1/a_{\rm
fut}$ (lower panel) for 119~considered LPCs. The uncertainties of $1/a$-determinations were
incorporated into these $1/a$-histograms by taking the full cloud of
VCs for each comet. This means that both distributions are composed of 119~individual
normalized 1/a-distributions resulting from the dynamical calculations of 5001\,VCs.
}
\label{fig:OortSpike_105}
\end{figure}

To derive the original and future inverse semi-major axis, the dynamical calculations 
of a swarm of VCs of each individual objects 
were performed backwards and forwards in time until it reached 
250\,au from the Sun, that is, at a distance where the planetary perturbations are already completely negligible. 
Each individual swarm of starting osculating orbits consists of
5\,001\,VCs including the nominal orbit and was constructed according
to a Monte Carlo method proposed by \citet{sitarski:1998}, where the
entire swarm fulfilled the Gaussian statistics of fitting to positional
data used for a given osculating (nominal) orbit determination. This method allowed us
to determine the uncertainties of original and future orbital elements,
including the inverse semi-major axes by fitting each orbital element of original/future 
swarm of VCs to Gaussian distribution 
(see Tables~\ref{tab:orbit_original_27}--\ref{tab:orbit_future_59}).

The distribution of $1/a_{\rm ori}$ based on the best solutions derived from past dynamical
studies and incorporating the uncertainties of $1/a_{\rm ori}$-determinations is shown in
Fig.~\ref{fig:OortSpike_105}. The only object from the sample of 119~comets
seems to be a serious candidate to interstellar comet (C/2007~W1 Boattini, for more details see Paper~4).
The left and negative wing of the main distribution is formed by a few comets with rather broad individual $1/a_{\rm ori}$-distributions (Gaussians),
such as C/1978~G2 (see Table~\ref{tab:orbit_original_59}) with a formally negative, but poorly known, inverse semi-major axis of $-22.4\pm 37.8$ in units of $10^{-6}$\,au$^{-1}$, or by C/1984~W2 with $1/a_{\rm ori} = 36.1\pm 20.2$ in the same units. 
The main peak of $1/a_{\rm ori}$ is broad and extends
between 10 and 65 in units of $10^{-6}$\,au$^{-1}$ ($a_{\rm ori}$: $\sim$15\,400 -- 100\,000 au).
More precisely, the calculated 10 and 90 per cent deciles are 11.5 and 74.5 in units of $10^{-6}$\,au$^{-1}$, respectively. 
The median value of the $1/a_{\rm ori}$-distribution is $37.95\cdot 10^{-6}$\,au$^{-1}$ ($a_{\rm ori}\simeq$26\,300\,au).
The main part of $1/a_{\rm ori}$-distribution seems to have a local minimum somewhere between 25--35 in units of $10^{-6}$\,au$^{-1}$, but the histogram based on twice wider bins gives a broad single maximum. 
This dip in broad Oort maximum is exclusively caused by large-perihelion comets. 
Thus, significantly richer statistics are needed to confirm or refute the double nature of this maximum.

\subsection{Part~IV of the catalogue. Future orbital elements}\label{sec_future_orbits108}

This part consists of Tables~\ref{tab:orbit_future_27}, \ref{tab:orbit_future_22} and \ref{tab:orbit_future_59} (samples A1, A2, B) giving future barycentric orbits calculated at a distance of 250\,au from the Sun. 

\ni The $1/a_{\rm fut}$-distribution constructed from the future orbits presented in 
these tables are shown in the lower panel of Fig.~\ref{fig:OortSpike_105}. Fifteen~comets ($\sim$13\,per cent of the sample) have a future inverse semi-major axis within the same range of $0<1/a_{\rm fut}<100\cdot 10^{-6}$\,au$^{-1}$ where the original 1/a-distribution is placed, $\sim$51\,per cent of comets leave the solar system on hyperbolic orbits, and about 36\,per cent of objects have evolved into more tightly bound orbits under the influence of planetary perturbations.
The narrow peak in the range $|1/a_{\rm fut}|<100\cdot 10^{-6}$\,au$^{-1}$ is evident and includes more than 30 per cent of the entire sample. This peak mostly consists of comets with large perihelion distances ($q_{\rm osc}>3.1$\,au), which means that these comets reproduced 78.5 per cent of the peak, whereas the entire sample of 118 objects taken for this statistics (comet C/2010~X1 Elenin disintegrated shortly after perihelion passage)
includes only 69 large-perihelion comets (58 per cent of the sample).

\section{Concluding remarks} \label{sec_conclusions}

\subsection{About accuracy} \label{subsec_about_accuracy}

\begin{figure}
\includegraphics[width=8.8cm]{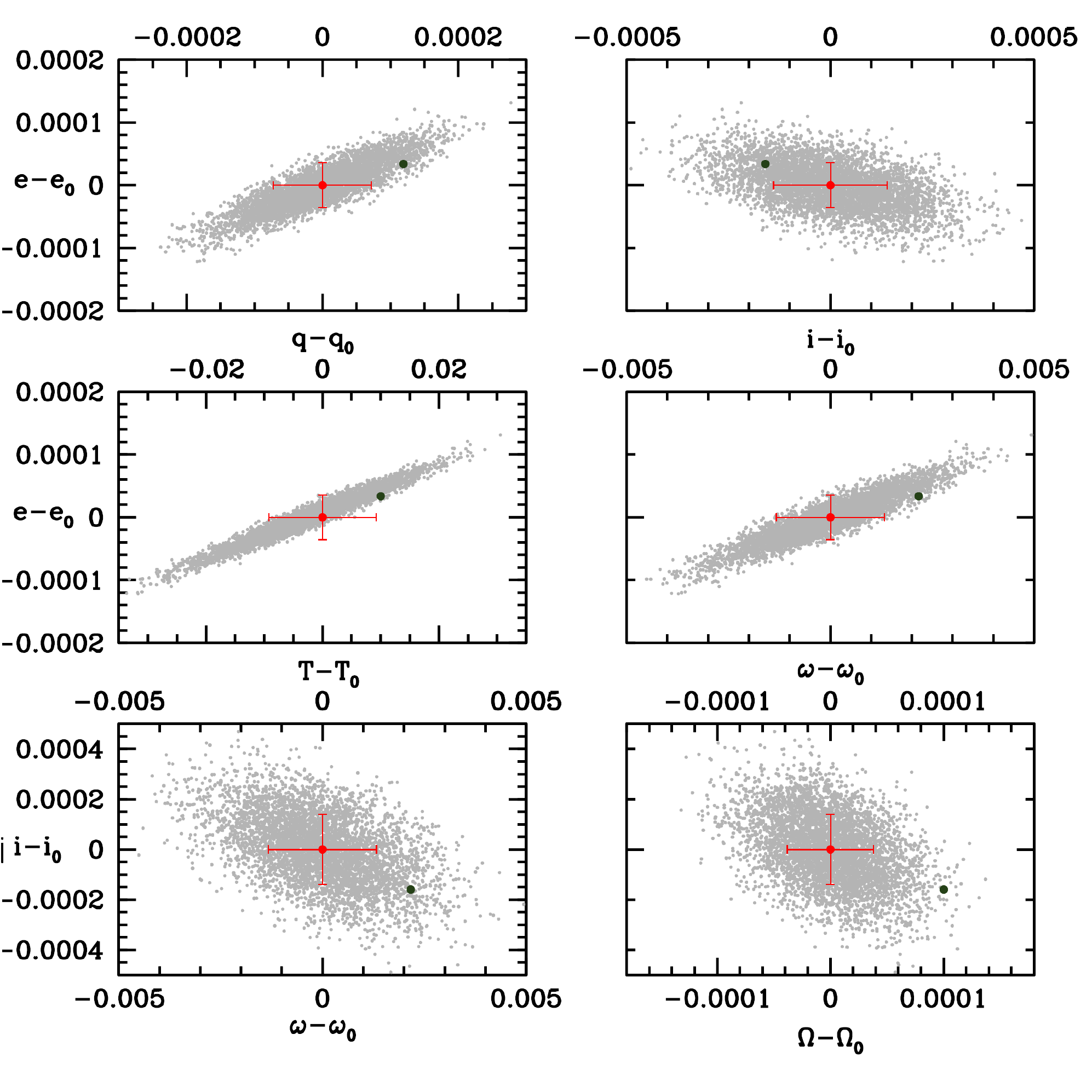}
\caption{Projection of the 6D space of 5\,001~VCs of C/1988~B1 onto six chosen planes of osculating orbital elements. 
Vertical axes given in the right-hand panels are exactly the same as the vertical axes in the left-hand panels. 
Each grey point represents a single virtual orbit, while the red large 
point represents the nominal orbit solution given here and the green point is taken from MWC\,08
(see also Table~\ref{tab:orbit_comparison}). Each plot is centred on the nominal values of the respective pair of osculating orbital elements denoted by the subscript '0'; see also Table~\ref{tab:orbit_comparison}.}
\label{fig:1988b1_VCs}
\end{figure}

\begin{table*}
\caption{\label{tab:orbit_comparison} Heliocentric orbital elements of osculating orbits for comets C/1988~B1 Shoemaker and C/2002~J4 NEAT (Sample~B, see also Table~\ref{tab:orbital_quality108_list}). 
The successive Cols. signify $[2]$ -- epoch, i.e. osculation date,
$[3]$ -- perihelion time [TT], $[4]$ -- perihelion distance, $[5]$
-- eccentricity, $[6]$ -- argument of perihelion (in degrees),
equinox 2000.0, $[7]$ -- longitude of the ascending node (in
degrees), equinox 2000.0, $[8]$ -- inclination (in degrees), equinox
2000.0. In the last line we present for each comet the difference between the nominal and catalogue values of a given element divided by the respective combined error, thus Diff$_{\rm i} = {\rm elem}_{\bf nom,i}-{\rm elem}_{\bf cat,i} /(\sqrt{2}\delta _i)$.}
\center{\tiny{
{\setlength{\tabcolsep}{4.3pt}
\begin{tabular}{ccrrrrrrrl}
\hline \hline 
           & Epoch     & T$_0$            & q$_0$         &  e$_0$        & $\omega _0$ & $\Omega _0$ &  i$_0$  & $1/a_{\rm ori,0}$ & Refferences\\
\hline \\
\multicolumn{10}{c}{Comet C/1988 B1 Shoemaker} \\
 elem$_{\rm nom,i}$ &  19870326 &  19870320.107555 &    5.03070649 &    1.00248504 &  124.217043 &  325.159660 &   80.585940  &    20.0   & here   \\
 $\delta _i $       &           &     $\pm$.009206 &$\pm$.00007222 &$\pm$.00003544 &$\pm$.001331 &$\pm$.000038 &$\pm$.000139  & $\pm$7.0  &        \\
 elem$_{\rm cat,i}$ &  19870326 &  19870320.12030~~~&    5.0308257~~~&    1.0025188~~~&  124.21921~~~&  325.15976~~~&   80.58578~~~& 13~~~ & MWC\,08\\ 
Diff$_{\rm i}$ & & 0.98~~~~~~~~ & 1.17~~~~~~~~~~~~ & 0.67~~~~~~~~~~~~ & 1.15~~~~~~~~ & 1.86~~~~~~~~ & 0.81~~~~~~~~ &  0.71\hspace{-0.12cm} & \\
           \\
\multicolumn{10}{c}{Comet C/2002 J4 NEAT} \\
 elem$_{\rm nom,i}$ &  20031008 &  20031003.151072 &    3.63378023 &    1.00001831 &  230.705722 &   70.881227 &   46.521834  &    33.9   & here\\
 $\delta _i $       &           &     $\pm$.000235 &$\pm$.00000105 &$\pm$.00000356 &$\pm$.000035 &$\pm$.000014 &$\pm$.000015  & $\pm$1.0  & \\
 elem$_{\rm cat,i}$ &  20031008 &  20031003.15132~~~&    3.6337804~~~&    1.0000199~~~&  230.70575~~~&   70.88122~~~&   46.52183~~~& 35~~~ & MPC 75513\\  
Diff$_{\rm i}$  & & 0.75~~~~~~~~ & 0.11~~~~~~~~~~~~ & 0.45~~~~~~~~~~~~ & 0.57~~~~~~~~ & 0.35~~~~~~~~ & 0.19~~~~~~~~ & 0.78\hspace{-0.12cm} & \\
\hline
\end{tabular}}}}
\end{table*}

In this catalogue great care was taken to treat the data completely homogeneously together with an individual approach to determining NG~effects wherever it seemed necessary. In Papers\,1--2 it was shown that data treatment is crucial for an orbit determination of one-apparition comets and the applied methods of data selection and weighting are there discussed in details. 
An individual approach was applied when there were any doubts about the O-C~time variation of residuals or the O-C~distribution of residuals (for the standard model based on the all available data). Thus, sometimes two sets of osculating orbits are offered for a given comet, in particular, for some comets with determinable NG~effects. In these cases, the first set always gives the overall model of NG~motion derived using the entire set of data. The second set describes the orbit derived using a subset of data, e.g. using only the pre-perihelion data for ingoing leg of orbit, and post-perihelion data for outgoing leg of orbit, respectively (see for example the case of C/2007~T2 or C/2008~A1). This approach was necessary because of the simplicity of the NG~model used for orbital calculations (usually three constant NG~parameters, see Table~\ref{tab:NG-parameters}). In fact, it was the only possible model for the purpose of such a massive NG~determination in the motion of LPCs. A detailed discussion of the different forms of NG~accelerations is given in Paper~3 for the two small-perihelion comets C/2001~Q4 and C/2007~T2. In that paper, we showed that  determining the exponents for the $g(r)$-like function is very difficult and, in practice, this was only possible for these two particular comets from the catalogue sample. 

\noindent An additional note is needed for large-perihelion comets with NG~orbits in the catalogue. The $g(r)$-form of NG~accelerations used here describes the water ice sublimation. Thus, this relationship is very coarse for comets with large perihelion distances, where the percentage of the more volatile ices may be relatively high.

In the context of this discussion, we strongly emphasize that all the uncertainties of osculating orbital elements given in the catalogue tables (and hence the uncertainties of original and future orbits) represent the formal errors derived by fitting the assumed model of motion (GR or NG) to positional observations. If it were possible to explore the NG~acceleration along the orbit of a given comet more precisely, the solution might be different. How different depends on the case in question. However, as was shown in Paper~3, the scatter of $1/a_{\rm ori}$-values for various NG~models appears to be significantly smaller than the difference between the $1/a$-values between the pure GR and the standard NG~model for the two comets examined there. Therefore, it seems that this relatively simple type of model of NG~acceleration applied for the purpose of this catalogue gives orbits that are significantly closer to the actual past and future motion of analysed comets. When the various NG~orbits are given in the catalogue, a more cautious approach to the formal errors is necessary. 
For this reason, in the case of comet C/2007~W1  Boattini, which is the best candidate for a comet on an original hyperbolic orbit, the various NG~models were taken into account (see Paper~4) to be conclusive about the shape of its barycentric original orbits. We concluded in Paper~4 that all of them indicate that this comet moved on a barycentric hyperbolic orbit before entered the planetary zone. This makes this comet the best candidate to have an interstellar origin among the catalogue comets. 

Generally, the GR~orbits given in this catalogue differ from the solutions given in MWC\,08 or \citet{IAU_MPC_Web} by no more than $3\sigma$-error for each orbital element; typically by even less than a $2\sigma$-error. Two representative examples are given in Table~\ref{tab:orbit_comparison} for comets C/1988~B1 and C/2002~J4, which have 1a and 1a+ quality class orbit, respectively. Both solutions are based on the data span over long time intervals (more than 4 yr). 
In the last line for each comet in Table~\ref{tab:orbit_comparison} the difference in each orbital element is expressed as combined errors and assuming the same orbital element uncertainties for the solution given in MWC\,08 \citep{IAU_MPC_Web} as derived here. Additionally, for C/1988~B1, the solution taken from MWC\,08 is compared in Fig.~\ref{fig:1988b1_VCs} with the swarm of osculating VCs derived for the past and future evolutionary calculations. Orbit of C/1988~B1 given in MWC\,08 was determined using 112 residuals (rms=1\farcs 00), whereas the orbit presented here was obtained on the basis of 127 residuals (rms=0\farcs 98, without weighting of the data). 

An extensive discussion of possible differences of the NG~solutions given in this catalogue from other published results is presented in Paper~4 for two comets with variable NG~effects, C/2007~W1 and C/2008~A1. 

For comets with two different NG~solutions given in this catalogue, both orbits can be taken by the user to investigate a cometary activity within the observational interval. However, for long-time orbital calculations, the more dedicated orbit is recommended, for example to  investigate the cometary origin, the orbit determined using the pre-perihelion data should be taken into account. 

The web~catalogue also offer the swarms of original and future VCs, that is, the barycentric orbital elements at the distances where planetary perturbations are negligible for all models from Table~\ref{tab:Comets108_threesamples}. Thus, the swarms of original and future NG~orbits are also available in this catalogue for further investigation by the user. 

\subsection{Future plans}

This catalogue will be supplemented in the future by the sample of one-apparition comets discovered in the first half of the twentieth century. Orbits of such old comets particularly need to be redetermined because they were originally obtained using very different numerical methods and assumptions on the model of the solar system, including the number of planets taken into account. The first part of this long-standing project describing the analysis of 38~comets from the Oort spike will be soon submitted to publication. 

One of the issues awaiting for extensive numerical tests is whether the gravitational influence of the mass of the 
asteroid belt, Kuiper Belt, and Oort Cloud will have a strong effect in particular cases on the orbital elements of near-parabolic comets that general move on steeply inclined orbits to the ecliptic plane. Some test were made for the most massive bodies from the Main Belt 
(which seems to be the more important agent for the comets investigated here), and no important orbital changes were so far noticed for a few exercised comets. The uncertainties of orbital elements were more than two orders of magnitude greater than the differences in nominal orbit determined using the standard solar system model and that taking into account four massive asteroids from the Main Belt. 
However, without more scrupulous testing one cannot rule out the rare possibility that a particular comet will be sensitive to perturbations coming from one of the relatively massive bodies (due to close encounter). In the future it will be potentially interesting to incorporate the many of massive minor planets into the dynamical model of the solar system and perform such tests for all comets with previously determined orbits. Additionally, it is not obvious how the global mass distribution of mentioned three populations of solar system minor bodies are gravitationally important for near-parabolic comets. These plans of comprehensive tests will be gradually put into practice. 

%\newpage

\begin{acknowledgements}
I wish to thank the reviewer Luke Dones for many constructive suggestions that improved this paper. \newline 
The orbital calculation was performed using the numerical orbital package developed by Professor Grzegorz Sitarski and the Solar System Dynamics and Planetology Group at SRC PAS. 
\end{acknowledgements}
\bibliographystyle{aa}
%\bibliography{moja22app,mkr_22comets}

{\onecolumn{

\appendix

%%%%%%%%%%%%%%%%%%%%%%%%%%%%%%%%%%%%%%%%%%%%%%%%%%%%%%%%%%%%%%%%%%%%%%%%%%%%%%%%%%
%\setcounter{table}{6}
%%%%%%%%%%%%%%%%%%%%%%%%%%%%%%%%%%%%%%%%%%%%%%%%%%%%%%%%%%%%%%%%%%%%%%%%%%%%%%%%%%

%\begin{sidewaystable}[h]
\begin{landscape}
\section{Part\,I -- Description of observational material and orbital quality assessment}

% \resizebox{\textheight}{!}{  }
%\begin{center}
%\begin{savenotes}

%\begin{longtab}
\setlength\LTcapwidth{1.35\textwidth} % default: 4in (rather less than \textwidth...)
\setlength\LTleft{0pt}               % default: \parindent
\setlength\LTright{0pt}              % default: \fill
{\setlength{\tabcolsep}{1.8pt}{
\begin{longtable}{lcccrrcccccccc}
%{\footnotesize {
\caption{\label{tab:orbital_quality108_list}
%\vspace{0.1cm}
Characteristics of observational material and
the new quality assessment for 119~near-parabolic comets investigated in
Papers~1--4 (see references in the last col.). The sample includes
comets with first-quality orbits or NG~orbits in MWC\,08 and
those discovered during the years 2006--2010. The observational
material of comets taken for osculating orbit determination is
described in Cols.\,$[2]-[8]$ and $[13]$, whereas the new orbit
quality assessment resulting from the osculating orbit based on these
data sets is shown in Cols.\,$[10]-[12]$. 
The second and third cols. show an osculating
perihelion distance and perihelion time. The data
distribution relative to a perihelion passage is presented in
Cols.\,$[7]$ \& $[8]$, where 'pre!' ('post!') means that all
observations were taken before (after) perihelion passage; 'pre+'
('post+') means that significantly more pre-perihelion
(or post-perihelion) measurements were available, and an additional '$+$' indicates
the drastic dominance of pre-perihelion (or post-perihelion)
measurements in both the number and the time interval. Col.\,$[9]$
shows the type of the best model possible to determine from the full
interval of data, where the subscript 'un' indicates that 
the GR and even the NG~model determined from the entire
interval of data is not satisfactory in the sense of the three
criteria described in detail in Papers~1--4. For such an
unsatisfactory orbital solution the better model for pre-perihelion
motion (that based on some subsample of data) is given in the second
row of a given comet, except for comets C/1990~K1 and C/1993~A1,
where the individual data treatment is has not been yet.
Cols.\,$[10]$ and $[11]$ give the Q$^*$ for GR and NG~models,
respectively. The orbital class resulting from the model described
in Col.\,$[9]$ and determined according to the new quality
assessment is given in Col.\,$[12]$. References to publications where dynamical studies were based on the orbital solution presented here are listed in the last col., where P1
refers to \citet{kroli-dyb:2010}, P2 to \citet{dyb-kroli:2011}, P3 to \citet{kroli-dyb:2012}, and P4 to \citet{kroli-dyb:2013}. Many additional notes to individual objects  are appended to this table, and the remark 'new' indicates solutions that have not been used in P1-P4.} \\
%\begin{tabular}
\hline\hline
%\toprule
Comet & q$_{\rm osc}$ & T            & Observational arc        & No  & Data      & Heliocentric   & Data  & Type & \multicolumn{2}{c}{Q$^*$} & New     & rms [arcsec]  & Ref  \\
name  &            &                 & dates                    & of  & arc span  & distance span  & type  & of   &   GR    & NG              & orbital & /no of &      \\
      & {[}au{]}   & {[}yyyymmdd{]}  & {[}yyyymmdd--yyyymmdd{]} & obs & {[}yr{]}  & {[}au{]}       &       & model&   model & model           & class   & res.  &      \\
$[1]$ & $[2]$  & $[3]$     & $[4]$              & $[5]$ & $[6]$& $[7]$     & $[8]$& $[9]$&$[10]$&$[11]$&$[12]$& $[13]$& $[14]$\\
\hline
\endfirsthead
\caption{continued.}\\
\hline\hline
%
%\midrule
%\endfirsthead
%
%\multicolumn{14}{c}%
%{{\bfseries \tablename\ \thetable{} -- continued from the previous page}}\\
%\toprule
Comet & q$_{osc}$  & T               & Observational arc        & No  & Data      & Heliocentric   & Data  & Type & \multicolumn{2}{c}{Q$^*$} & New     & rms [arcsec]  & Ref  \\
name  &            &                 & dates                    & of  & arc span  & distance span  & type  & of   &   GR    & NG              & orbital & /no of & \&   \\
      & {[}au{]}   & {[}yyyymmdd{]}  & {[}yyyymmdd--yyyymmdd{]} & obs & {[}yr{]}  & {[}au{]}       &       & model&   model & model           & class   & res.  & Notes \\
$[1]$                            & $[2]$  & $[3]$     & $[4]$              & $[5]$ & $[6]$& $[7]$     & $[8]$& $[9]$&$[10]$&$[11]$&$[12]$& $[13]$& $[14]$\\
\hline
\endhead
\hline 
\endfoot
%
%\bottomrule
%\endlastfoot
%
%%
\\
\multicolumn{14}{c}{\bf Sample A1. Years 1885--2005. 28 near-parabolic comets of $q_{\rm osc}<3.1$\,au $1/a_{\rm ori}<150\times 10^{-6}$\,au$^{-1}$} \\
C/1885 X1   Fabry                & 0.642  & 18860406  & 18851201--18860719  & 228  & .63  & 2.29--2.00  & full & NG  & 6.5 & 6.0 & 2a & 3.58/ 390& P1,1  \\
C/1892 Q1   Brooks               & 0.976  & 18921228  & 18920901--18930713  & 191  & .85  & 2.11--3.04  & full & NG  & 6.5 & 6.0 & 2a & 2.80/ 334& P1,1  \\
C/1952 W1   Mrkos                & 0.778  & 19530124  & 19521210--19530718  & 36   & .60  & 1.16--2.86  & full & NG  & 6.0 & 5.5 & 2a & 1.08/  61& P1,2  \\
C/1956 R1   Arend-Roland         & 0.316  & 19570408  & 19561108--19580411  & 249  & 1.4  & 2.83--5.36  & full & NG  & 7.0 & 7.0 & 1b & 1.39/ 458& P1,1  \\
C/1959 Y1   Burnham              & 0.504  & 19600320  & 19600104--19600617  & 88   & .45  & 1.63--1.81  & full & NG  & 6.0 & 5.0 & 2b & 1.60/ 146& P1,1  \\
C/1974 F1   Lovas                & 3.011  & 19750822  & 19740321--19770911  & 137  & 3.5  & 5.64--7.26  & full & NG  & 8.0 & 8.0 & 1a & 1.08/ 273& P2  \\
\end{longtable}
\end{longtable}
}}
\tablefoot{Notes in column $[14]$: 1 -- Many more observations were used here than were used for the MWC\,08 orbit or the orbit given at the \citet{IAU_MPC_Web} (as of May 2013), 2 -- Significantly fewer observations of C/1952~W1 was available for us than were used for the MWC\,08 orbit determination.\\
\tablefoottext{3}{This new solution is based on a three days longer data interval and 4 more observations than the orbit analysed in Paper~2.}
\tablefoottext{4}{Orbit is based on a $\sim$6.5 longer data interval and 3 more observations than that analysed in Paper~2.}
\tablefoottext{5}{Orbit is based on a $\sim3\,$weeks longer data interval and 4 more observations than that analysed in Paper~2.}
\tablefoottext{6}{Orbit is based on a $\sim1.6\,$yr longer data interval and 4 more observations than that analysed in Paper~2; this results in an increase from quality class 1a to 1a+.}
\tablefoottext{7}{Orbit is based on a $\sim$two month longer data interval and 9 more observations than that analysed in Paper~2.}
\tablefoottext{8}{Orbit is based on a $\sim$two month longer data interval and 2 more observations than that analysed in Paper~2.}}
%
%}}}}
%\end{longtable}

\end{landscape}

\centerline{{\bf Additional notes on some individual objects given
in Table~\ref{tab:orbital_quality108_list}}}

\vspace{0.2cm}

\noindent \textbf{Comet C/1990~K1}. MWC\,08 gives an NG~orbit based on
significantly fewer observations (314) and the same time interval as
in the table. In June 2013, 693 observations were available at the
\citet{IAU_MPC_Web}, but the osculating orbit was the same as in
MWC\,08. Since MWC\,08 gives an NG~orbit, orbital quality class is
not given there. The pure GR osculating orbit (1a quality
class\footnote{At \citet{JPL_Browser} no orbital quality assessment
is published, thus we calculated here orbital class using our new
scheme}) given at the \citet{JPL_Browser} is based on 553 observations.
In Paper~1 the symmetric NG~model based on full data interval was
analysed, where some trends in O-C variations were reported. The present
symmetric NG~solution results in 1a quality class. The asymmetric
model (function ${\rm g}(r(t-\tau))$) listed in this catalogue
gives $\tau = 5.3\pm 1.7$\,day and orbit of class~1b. However, still some trends in O-C time
variations are easily seen and a more dedicated treatment is necessary
for this comet, maybe similar to that in Paper~3 for C/2001~Q4 and
C/2002~T7.

\vspace{0.2cm}

\noindent \textbf{Comet C/1993~A1}. MWC\,08 gives NG~orbit based on
a poorer data set (539 observations) and shorter time interval
(1993 January~2 -- 1994 June~10). In June 2013, 745 observations were
available at the \citet{IAU_MPC_Web}, but the osculating orbit was the
same as in MWC\,08. Since MWC\,08 gives an NG~orbit, an orbital
quality class is not given there. In Paper~1 the symmetric NG~model
based on the full data interval was used (orbit of class~1a) and some
trends in O-C variations were reported. The asymmetric model
(function ${\rm g}(r(t-\tau))$)  presented here gives $\tau = -34.2\pm 3.2$\,day and
slightly decreasing rms (orbit of class~1a) compared with the GR~model,
but still some trends in O-C time variations are easily
recognizable. Probably a more dedicated treatment is necessary for
this comet, maybe similar to that in Paper~3 for C/2001~Q4 and C/2002~T7.

\vspace{0.2cm}

\noindent \textbf{Comet C/2001 Q4}. MWC\,08 gives an NG~orbit based on
significantly fewer observations (1106) and shorter arc (2001 August~4
-- 2004 June~11). In June 2013, 2681 observations were available at
the \citet{IAU_MPC_Web}, but the osculating orbit was the same as in
MWC\,08. Since MWC\,08 gives an NG~orbit, the orbital quality class is
not given there. The NG~osculating orbit (quality class~1a+) given at
the \citet{JPL_Browser} is based on 2567 observations. Here, two models
are listed: (1) an NG~solution based on a full data interval and (2) the
most recommended model for past and future dynamical evolution.
The first model gives NG~orbit of class~1a+ with evident trends in O-C
time variations. The asymmetric model gives $\tau = 0.73\pm 0.45$\,day,
and no decrease in rms, therefore the symmetric model is included in this catalogue.
The second model is based on data taken at large
distances and gives orbit of class~1a+ and O-C time variations
with no trends in right ascension and declination. In Paper~3, we recommend
this model for dynamical studies and investigation of the origin of this
comet. For more details see model DIST2 in Paper~3.

\vspace{0.2cm}

\noindent \textbf{Comet C/2002 T7}. MWC\,08 gives an NG~orbit based on
3825 observations and a significantly shorter arc (2002 October~12 --
2004 June~11). In June 2013, 4517 observations were available at
the \citet{IAU_MPC_Web}, which span the same time interval as in this
table. However, the osculating orbit was the same as in MWC\,08. Since
MWC\,08 gives an NG~orbit, the orbital quality class is not given
there. The NG~osculating orbit (quality class~1a+) given at
the \citet{JPL_Browser} is based on 4399 observations and a slightly
shorter data interval (2002 October~12 -- 2006 March~07). Here, two models
are listed: (1) an NG~solution based on full data interval and (2) the
most recommended models for past and future dynamical evolution, respectively.
The first model gives an NG~orbit of class~1a+ with evident trends in O-C
time variations. The asymmetric model gives $\tau = -12.9\pm 0.6$\,day,
$|{\rm A}_1| <<|{\rm A}_2|$ and an infinitesimal decrease in rms, 
therefore the symmetric model is shown in this catalogue. The second (PRE) and third (POST) 
models are based on data taken before and after
perihelion passage, respectively, and give orbits of class~1a and 1b as well as O-C time
variations with no trends in right ascension and declination. In Paper~3, we
recommend osculating orbits based on these two models to study the origin and future 
of this comet. For more details see model PRE in Paper~3.

\vspace{0.2cm}

\noindent \textbf{Comet C/2003~K4}. MWC\,08 gives an GR~orbit based on
significantly fewer observations (2244) and shorter arc (2006 May~28
-- 2004 July~09). In June 2013, 3712 observations were available at
the \citet{IAU_MPC_Web} that span the same time interval as in this
table, but the orbit was the same as in MWC\,08. The NG~osculating orbit 
(quality class~1a+) given at the \citet{JPL_Browser} is based on 3606
observations. In Paper~1 the symmetric NG~model based on a full data
interval is given (orbit of class~1a+) and some trends in O-C
variations are reported. The asymmetric model (function ${\rm
g}(r(t-\tau))$) gives a very long time shift $\tau = -89\pm 3$\,day and slightly 
decreasing rms (orbit of class~1a) compared with the symmetric
NG~solution, but still similar trends in O-C time variations are
easily visible. Thus, two models
are listed here: (1) the symmetric NG~solution based on the full data interval 
(where observations are weighted slightly differently from the solutions 
investigated in Paper~1), and (2) the most recommended model for past and future 
dynamical evolution based on observations taken 
at large heliocentric distance from the Sun. In the second model no trends in O-C
time variations are noted.

\vspace{0.2cm}

\noindent \textbf{Comets C/2006~HW$_{51}$, C/2006~L2 and C/2008 T2}.
In Paper~4, we decided to analyse the GR~solutions since the NG~solutions give a negative normal component of NG~acceleration (parameter A$_1$ < 0) for these comets which makes
the NG~solution rather uncertain. For more details see Paper~4.

\vspace{0.2cm}

\noindent \textbf{Comet C/2006 VZ$_{13}$}. MWC\,08 gives a GR~orbit
(class~1B) based on 356 observations and the time interval of 2006 November~13
-- 2007 June~27. In June 2013, 1173 observations were available at
the \citet{IAU_MPC_Web} span the same time interval as the first model for this comet 
in our table. However, the orbit given at the \citet{IAU_MPC_Web} was the same as
in MWC\,08. The GR~osculating orbit (quality class~1b) given at
\citet{JPL_Browser} is based on 1037 observations and the time interval
2006 November~13 -- 2007 August~05. Since observations have stopped soon after
perihelion passage, a dedicated solution is only possible for
pre-perihelion data. Here, two models are listed: (1) the NG~solution based on the full data interval, and (2) the most recommended model for past dynamical evolution. 
In Paper~4, we recommend this second solution (of quality class~2b) for backward dynamical studies and a NG~solution based on all data for forward extrapolation of motion (class~1b) with the remark that the future motion of this comet is additionally uncertain because
of the lack of post-perihelion data. For more details see Paper~4.

\vspace{0.2cm}

\noindent \textbf{Comet C/2007 N3}. Evident trends in the
O-C time variations for the NG~solution (important normal component of NG~acceleration)
based on an entire data set  were noticed. Therefore we list in this catalogue
two types of models: (1) an asymmetric NG~solution based on a full data interval, and (2) the most recommended two separate GR~models based on pre-perihelion and post-perihelion data, for past and future dynamical evolution, respectively. Fore more details see Paper~4.

\vspace{0.2cm}

\noindent \textbf{Comet C/2007 O1}. According to the \citet{JPL_Browser},
this comet was also known as 2006~GA$_{38}$, therefore eight
positional observations of this object were also used for orbit determination.
For more details see Paper~4.

\vspace{0.2cm}

\noindent \textbf{Comet C/2007 Q1}. In MWC\,08 and the
\citet{JPL_Browser} only a parabolic orbit for this comet is given
(assumed e=1 thus unknown 1/a, no quality class). For more details see
Paper~4.

\vspace{0.2cm}

\noindent \textbf{Comet C/2007 Q3}. At the \citet{IAU_MPC_Web} over
ten times more observations are available than were used in MWC\,08 for
the orbit determination.  In Paper~4, we noted visible
trends in the O-C solution for an NG~solution based on all positional data.
Therefore we list two types of models: 
(1) a symmetric NG~solution based on a full data interval, and (2) the
most recommended two separate GR~models based on pre-perihelion and post-perihelion data, 
for past and future dynamical evolution, respectively. For more details see
Paper~4.

\vspace{0.2cm}

\noindent \textbf{Comet C/2007 W1}. MWC\,08 includes only a pure GR
orbit of a class~1B for this comet (based on 344 observations taken
until Mar 16, 2008). Here, 1703 observations given at the
\citet{IAU_MPC_Web} were taken. Some part of them were used at
the \citet{IAU_MPC_Web} to recalculate an orbit of this comet using the NG~model. The GR~osculating orbit (quality class~1b) given at the \citet{JPL_Browser} is based on
599 observations and the data interval of 2007 November~20 -- 2008 August~30. 
A~full asymmetric NG~model gives orbit of class~1a, but very strong trends in
O-C diagram were noted and NG~effects seem to be variable in the
motion of C/2007 W1. Therefore in Paper~4, similarly to \citet{NK1731A,NK1731B}, we
determined two dedicated NG~orbits for backward and forward
dynamical studies. The orbit based on pre-perihelion data is of
class~1b while the post-perihelion solution gives an orbit of
quality class~2b. In this catalogue the asymmetric NG~solution based on the entire data set 
is slightly different from that used in Paper~4 (observations were selected using a more
restrictive criterion). A detailed discussion of this unique comet is given in Paper~4.

\vspace{0.2cm}

\noindent \textbf{Comet C/2008~A1}. MWC\,08 includes only a pure GR
orbit of a class~1B for this comet (basing on 240 observations taken
until May 15, 2008). Here, 937 observations given at
the \citet{IAU_MPC_Web} were used. Some parts of them (until May 1, 2009) were used
at the \citet{IAU_MPC_Web} to recalculate NG~orbit of this comet. 
The GR~osculating orbit (quality class~1a) given at \citet{JPL_Browser} is based on 595
observations and the data interval of 2008 January~10 -- 2010 January~17. See also
\citet{NK1807}. The full asymmetric NG~model gives orbit of class~1a, but
very strong trends in the O-C diagram were easily seen and NG~effects
seem to be variable in its motion. Because of the erratic behaviour of
this comet we proposed in Paper~4 two different NG~solutions (both class~1b)
based on pre- and post-perihelion data (similarly to C/2007~W1).
More details are given in Paper~4.

\vspace{0.2cm}

\noindent \textbf{Comet C/2009~K5}. In June 2013, 2544
observations were available at the \citet{IAU_MPC_Web}, and the GR~osculating
orbit (class~1A) given there was based on 2307 observations, whereas
the GR~orbit (class~1a+) given at the \citet{JPL_Browser} was based on 2487
observations. Some trends in O-C time variations in the
pure GR~model (class~1a+) based on an entire data set were noted. Therefore we
recommend in Paper~4 two separate orbital solutions for the backward and
forward dynamical orbital evolution of this comet, both are pure GR
and give osculating orbits of class~1a. For more details see Paper~4.

\vspace{0.2cm}

\noindent \textbf{Comet C/2010~X1}. In June 2013, 2276
observations were available at the \citet{IAU_MPC_Web}, the GR~osculating
orbit (class~1A) given there was based on 1896 observations and on the time
interval of 2010 December~10 -- 2011 September~07. The GR~osculating orbit 
(quality class~1b) given at the \citet{JPL_Browser} is based on 2209
observations taken within the data interval of 2010 December~10 -- 2011 August~1.
This comet was observed until September~7, but started to
disintegrating in August. Therefore the data for GR~orbit determination
were taken here until the end of July for the first model (as at JPL),
and to the end of May for the second model. The first solution exhibits
some trends in O-C time variation, the second model is much better and
also gives orbit of class~1b. For more details see Paper~4.

\newpage

%\vspace{7.0cm}

\section{Part\,II -- Osculating orbital elements (heliocentric)}

%%%%%%%%%%%%%%%%%%%%%%%%%%%%%%%%%%%%%%%%%%%%%%%%%%%%%%%%%%%%%%%%%%%%%%%%%%%%%%
%	comets  with q < 3.1 au;  28 objects, osculating orbits
%%%%%%%%%%%%%%%%%%%%%%%%%%%%%%%%%%%%%%%%%%%%%%%%%%%%%%%%%%%%%%%%%%%%%%%%%%%%%%

%\begin{longtab}
\setlength\LTcapwidth{1.00\textwidth} % default: 4in (rather less than \textwidth...)
{\setlength{\tabcolsep}{2.5pt}{
\begin{longtable}{ccrrrrrrr}
\caption{\label{tab:orbit_osculating_27} Heliocentric orbital elements of osculating orbits for 28 near-parabolic comets described as sample~A1 in Table~\ref{tab:orbital_quality108_list}. 
The successive columns signify
$[1]$ -- comet designation, $[2]$ -- epoch, i.e. osculation date,
$[3]$ -- perihelion time [TT], $[4]$ -- perihelion distance, $[5]$
-- eccentricity, $[6]$ -- argument of perihelion (in degrees),
equinox 2000.0, $[7]$ -- longitude of the ascending node (in
degrees), equinox 2000.0, $[8]$ -- inclination (in degrees), equinox
2000.0,    $[9]$ -- inverse semi-major axis in units of $10^{-6}$\,au$^{-1}$. }\\
\hline\hline
%\toprule
Comet      & Epoch         & T                   & q            &  e         & $\omega$      & $\Omega$    &  i         & $1/a_{\rm osc}$ \\
           & $[$yyyymmdd$]$&$[$yyyymmdd.dddddd$]$& $[$au$]$     &            &  $[$\degr$]$  & $[$\degr$]$ & $[$\degr$]$& $[10^{-6}$au$^{-1}]$   \\
$[1]$      & $[2]$       & $[3]$                 & $[4]$        & $[5]$      & $[6]$         & $[7]$       & $[8]$      & $[9]$          \\
\hline
\endfirsthead
\caption{continued.}\\
\hline\hline
%\midrule
%\endfirsthead
%\multicolumn{9}{c}%
%{{\bfseries \tablename\ \thetable{} -- continued from the previous page}}\\
%\toprule
Comet      & Epoch         & T                   & q            &  e         & $\omega$      & $\Omega$    &  i         & $1/a_{\rm osc}$ \\
           & $[$yyyymmdd$]$&$[$yyyymmdd.dddddd$]$& $[$au$]$     &            &  $[$\degr$]$  & $[$\degr$]$ & $[$\degr$]$& $[10^{-6}$au$^{-1}]$   \\
$[1]$      & $[2]$       & $[3]$                 & $[4]$        & $[5]$      & $[6]$         & $[7]$       & $[8]$      & $[9]$          \\
\hline
\endhead
\hline
\endfoot
%\midrule
%\endhead
%\hline \multicolumn{9}{r}{{Continued on next page}} \\  \bottomrule
%\endfoot
%\bottomrule
%\endlastfoot
%%
 \\
\multicolumn{9}{c}{{\bf Osculating barycentric orbits based on the entire data sets}} \\
C/1885 X1  &  18860404 &  18860406.454383 &    0.64236680 &    1.00027315 &  126.588371 &   37.963059 &   82.628566 &  -425.22 \\
           &           &     $\pm$.000284 &$\pm$.00000206 &$\pm$.00002319 &$\pm$.000120 &$\pm$.000094 &$\pm$.000459 &$\pm$    36.10 \\
C/1892 Q1  &  18930117 &  18921228.586707 &    0.97595829 &    1.00030422 &  252.651332 &  266.034217 &   24.800982 &  -311.72 \\
           &           &     $\pm$.000228 &$\pm$.00000584 &$\pm$.00002555 &$\pm$.000427 &$\pm$.000181 &$\pm$.000152 &$\pm$    26.18 \\
%C/1913 Y1  &  19141104 &  19141026.767012 &    1.10444242 &    1.00009618 &   97.467000 &   60.396112 &   68.038602 &   -87.09 \\
%           &           &     $\pm$.000062 &$\pm$.00000218 &$\pm$.00000906 &$\pm$.000059 &$\pm$.000023 &$\pm$.000059 &$\pm$     8.20 \\
%C/1940 R2  &  19410106 &  19410116.234772 &    0.36774724 &    1.00040431 &  199.570995 &  296.590818 &   49.893845 & -1099.44 \\
%           &           &     $\pm$.000092 &$\pm$.00000232 &$\pm$.00001069 &$\pm$.000226 &$\pm$.000180 &$\pm$.000157 &$\pm$    29.06 \\
%C/1946 U1  &  19470224 &  19470207.363766 &    2.40762538 &    1.00086247 &  348.625215 &   35.558941 &  108.174289 &  -358.23 \\
%           &           &     $\pm$.000619 &$\pm$.00000738 &$\pm$.00001612 &$\pm$.000196 &$\pm$.000053 &$\pm$.000046 &$\pm$     6.69 \\
C/1952 W1  &  19530123 &  19530124.861090 &    0.77770934 &    1.00005338 &  253.822072 &  343.580012 &   97.186308 &   -68.64 \\
           &           &     $\pm$.000716 &$\pm$.00001008 &$\pm$.00014581 &$\pm$.000446 &$\pm$.000890 &$\pm$.000812 &$\pm$   187.49 \\
\bottomrule
\end{longtable}
}}
%}}
%\end{longtab}

%\newpage

%%%%%%%%%%%%%%%%%%%%%%%%%%%%%%%%%%%%%%%%%%%%%%%%%%%%%%%%%%%%%%%%%%%%%%%%%%%%%%
%	comets from Papers 4 with q < 3.1 au;  22 objects, osculating orbits
%%%%%%%%%%%%%%%%%%%%%%%%%%%%%%%%%%%%%%%%%%%%%%%%%%%%%%%%%%%%%%%%%%%%%%%%%%%%%%

%\begin{longtab}
\setlength\LTcapwidth{1.00\textwidth} % default: 4in (rather less than \textwidth...)
%\setlength\LTleft{0pt}               % default: \parindent
%\setlength\LTright{0pt}              % default: \fill
%{\footnotesize {
{\setlength{\tabcolsep}{2.5pt}{
\begin{longtable}{ccrrrrrrr}
\caption{\label{tab:orbit_osculating_22} Heliocentric orbital elements of osculating orbits for 22 near-parabolic comets described as sample~A2 in Table~\ref{tab:orbital_quality108_list}. 
The successive cols. signify
$[1]$ -- comet designation, $[2]$ -- epoch, i.e. osculation date,
$[3]$ -- perihelion time [TT], $[4]$ -- perihelion distance, $[5]$
-- eccentricity, $[6]$ -- argument of perihelion (in degrees),
equinox 2000.0, $[7]$ -- longitude of the ascending node (in
degrees), equinox 2000.0, $[8]$ -- inclination (in degrees), equinox
2000.0,    $[9]$ -- inverse semi-major axis in units of $10^{-6}$\,au$^{-1}$. }\\
\hline\hline
%\toprule
Comet      & Epoch         & T                   & q            &  e         & $\omega$      & $\Omega$    &  i         & $1/a_{\rm osc}$ \\
           & $[$yyyymmdd$]$&$[$yyyymmdd.dddddd$]$& $[$au$]$     &            &  $[$\degr$]$  & $[$\degr$]$ & $[$\degr$]$& $[10^{-6}$au$^{-1}]$   \\
$[1]$      & $[2]$       & $[3]$                 & $[4]$        & $[5]$      & $[6]$         & $[7]$       & $[8]$      & $[9]$          \\
\hline
\endfirsthead
\caption{continued.}\\
\hline\hline
%\midrule
%\endfirsthead
%\multicolumn{9}{c}%
%{{\bfseries \tablename\ \thetable{} -- continued from the previous page}}\\
%\toprule
Comet      & Epoch         & T                   & q            &  e         & $\omega$      & $\Omega$    &  i         & $1/a_{\rm osc}$ \\
           & $[$yyyymmdd$]$&$[$yyyymmdd.dddddd$]$& $[$au$]$     &            &  $[$\degr$]$  & $[$\degr$]$ & $[$\degr$]$& $[10^{-6}$au$^{-1}]$   \\
$[1]$      & $[2]$       & $[3]$                 & $[4]$        & $[5]$      & $[6]$         & $[7]$       & $[8]$      & $[9]$          \\
%\midrule
\hline
\endhead
\hline 
\endfoot
%\bottomrule
%\endlastfoot
\multicolumn{9}{c}{{\bf Osculating orbits from the entire data sets}} \\
C/2006 HW$_{51}$&20060922&20060929.310348 &    2.26564503 &    1.00224866 &  359.959683 &  228.127986 &   45.809385 &  -992.50 \\
           &           &     $\pm$.000161 &$\pm$.00000071 &$\pm$.00000774 &$\pm$.000044 &$\pm$.000038 &$\pm$.000024 &$\pm$     3.42 \\
C/2006 K3  &  20070301 &  20070313.363489 &    2.50143944 &    1.00085639 &  328.075919 &   49.402938 &   92.619710 &  -342.36 \\
           &           &     $\pm$.000363 &$\pm$.00000237 &$\pm$.00000599 &$\pm$.000136 &$\pm$.000029 &$\pm$.000020 &$\pm$     2.40 \\
C/2006 L2  &  20061101 &  20061120.199547 &    1.99394772 &    1.00093781 &   48.034772 &  239.246068 &  101.022297 &  -470.33 \\
           &           &     $\pm$.000122 &$\pm$.00000069 &$\pm$.00000284 &$\pm$.000043 &$\pm$.000012 &$\pm$.000023 &$\pm$     1.42 \\
C/2006 OF$_2$&20080911 &  20080915.684505 &    2.43143593 &    1.00083257 &   95.612164 &  318.508002 &   30.170411 &  -342.42 \\
\bottomrule
\end{longtable}
}}
%\end{longtab}

%\vfill\newpage

%%%%%%%%%%%%%%%%%%%%%%%%%%%%%%%%%%%%%%%%%%%%%%%%%%%%%%%%%%%%%%%%%%%%%%%%%%%%%%
%	comets from Papers 1 & 2 with q > 3.1 au;  59 objects osculating orbits
%%%%%%%%%%%%%%%%%%%%%%%%%%%%%%%%%%%%%%%%%%%%%%%%%%%%%%%%%%%%%%%%%%%%%%%%%%%%%%

%\begin{longtab}
\setlength\LTcapwidth{1.00\textwidth} % default: 4in (rather less than \textwidth...)
%\setlength\LTleft{0pt}               % default: \parindent
%\setlength\LTright{0pt}              % default: \fill
%{\footnotesize {
{\setlength{\tabcolsep}{2.5pt}{
\begin{longtable}{ccrrrrrrr}
\caption{\label{tab:orbit_osculating_59} Heliocentric orbital elements of osculating orbits for 69 near-parabolic comets described as sample~B in Table~\ref{tab:orbital_quality108_list}. All osculating orbits in this table are based on the entire data sets.
The successive cols. signify 
$[1]$ -- comet designation, $[2]$ -- epoch, i.e. osculation date,
$[3]$ -- perihelion time [TT], $[4]$ -- perihelion distance, $[5]$
-- eccentricity, $[6]$ -- argument of perihelion (in degrees),
equinox 2000.0, $[7]$ -- longitude of the ascending node (in
degrees), equinox 2000.0, $[8]$ -- inclination (in degrees), equinox
2000.0,    $[9]$ -- inverse semi-major axis in units of $10^{-6}$\,au$^{-1}$. }\\
\hline\hline
%\toprule
Comet      & Epoch         & T                   & q            &  e         & $\omega$      & $\Omega$    &  i         & $1/a_{\rm osc}$ \\
           & $[$yyyymmdd$]$&$[$yyyymmdd.dddddd$]$& $[$au$]$     &            &  $[$\degr$]$  & $[$\degr$]$ & $[$\degr$]$& $[10^{-6}$au$^{-1}]$   \\
$[1]$      & $[2]$       & $[3]$                 & $[4]$        & $[5]$      & $[6]$         & $[7]$       & $[8]$      & $[9]$          \\
\hline
\endfirsthead
\caption{continued.}\\
\hline\hline
%\midrule
%\endfirsthead
%\multicolumn{9}{c}%
%{{\bfseries \tablename\ \thetable{} -- continued from the previous page}}\\
%\toprule
Comet      & Epoch         & T                   & q            &  e         & $\omega$      & $\Omega$    &  i         & $1/a_{\rm osc}$ \\
           & $[$yyyymmdd$]$&$[$yyyymmdd.dddddd$]$& $[$au$]$     &            &  $[$\degr$]$  & $[$\degr$]$ & $[$\degr$]$& $[10^{-6}$au$^{-1}]$   \\
$[1]$      & $[2]$       & $[3]$                 & $[4]$        & $[5]$      & $[6]$         & $[7]$       & $[8]$      & $[9]$          \\
%\midrule
\hline
\endhead
\hline 
\endfoot
%\bottomrule
%\endlastfoot
%% 
\\
C/1972 L1  &  19721119 &  19721114.809745 &    4.27576201 &    1.00636268 &   56.657922 &  225.485976 &   79.367842 & -1488.08 \\
           &           &     $\pm$.003887 &$\pm$.00001790 &$\pm$.00002624 &$\pm$.000583 &$\pm$.000118 &$\pm$.000073 &$\pm$     6.13 \\
C/1973 W1  &  19730826 &  19730809.894467 &    3.84227589 &    1.00063440 &  221.260826 &  244.600667 &  108.066889 &  -165.11 \\
           &           &     $\pm$.006669 &$\pm$.00008706 &$\pm$.00004761 &$\pm$.002045 &$\pm$.000280 &$\pm$.000095 &$\pm$    12.39 \\
C/1974 V1  &  19740821 &  19740807.763518 &    6.01885058 &    1.00391195 &  151.758534 &  226.104055 &   60.853679 &  -649.95 \\
           &           &     $\pm$.018901 &$\pm$.00009141 &$\pm$.00007181 &$\pm$.001823 &$\pm$.000069 &$\pm$.000200 &$\pm$    11.90 \\       
C/1976 D2  &  19750128 &  19750115.546664 &    6.88076913 &    1.00210991 &  193.427853 &   22.786964 &  112.025871 &  -306.64 \\
\bottomrule
\end{longtable}
%}}
}}
%\end{longtab}

%\vfill\newpage

%%%%%%%%%%%%%%%%%%%%%%%%%%%%%%%%%%%%%%%%%%%%%%%%%%%%%%%%%%%%%%%%%%%%%%%%%%%%%%
%	NG~parameters for 45 comets with NG~solution in the Catalogue
%%%%%%%%%%%%%%%%%%%%%%%%%%%%%%%%%%%%%%%%%%%%%%%%%%%%%%%%%%%%%%%%%%%%%%%%%%%%%%

\begin{table*}
\caption{\label{tab:NG-parameters} NG~parameters derived in orbital solutions based on entire data intervals 
(first-line solution for a given comet marked STD in Col.\,$[6]$) and those based on a subsample of data marked  PRE  or  POST 
(see also Table~\ref{tab:orbital_quality108_list}). }
{\setlength{\tabcolsep}{9.5pt}
\begin{tabular}{@{}lc@{$\pm$}cc@{$\pm$}cc@{$\pm$}cc@{$\pm$}cc@{}}
\hline\hline
%\toprule
Comet    & \multicolumn{6}{c}{NG parameters defined by Eq.~\ref{eq:ng_std} in units of 10$^{-8}\,$au\,day$^{-2}$ } &\multicolumn{2}{c}{$\tau$} & model \\
         & \multicolumn{2}{c}{A$_1$}  & \multicolumn{2}{c}{A$_2$}  & \multicolumn{2}{c}{A$_3$} & \multicolumn{2}{c}{$[$ days$]$} & type \\
 $[1]$   & \multicolumn{2}{c}{$[2]$}  & \multicolumn{2}{c}{$[3]$}  & \multicolumn{2}{c}{$[4]$} & \multicolumn{2}{c}{$[5]$}    & {$[6]$} \\
\hline
\multicolumn{10}{c}{{\bf S a m p l e ~~~A1}} \\
C/1885 X1       & 2.337    & 0.295   &$-$0.288  & 0.161   & \multicolumn{2}{c}{--}   & \multicolumn{2}{c}{--}   &  STD   \\
C/1892 Q1       & 2.779    & 0.405   &   0.199  & 0.385   & \multicolumn{2}{c}{--}   & \multicolumn{2}{c}{--}   &  STD   \\
C/1952 W1       & 3.84     & 1.90    &$-$0.489  & 0.181   & \multicolumn{2}{c}{--}   & \multicolumn{2}{c}{--}   &  STD   \\
C/1956 R1       & 1.948    & 0.139   & 0.1123   & 0.0223  &   0.3728  & 0.0873       & \multicolumn{2}{c}{--}   &  STD   \\
C/1959 Y1       & 9.482    & 0.730   & 1.276    & 0.309   &   0.725   & 0.154        &\multicolumn{2}{c}{$-$9.3}&  STD   \\
C/1974 F1       & 50.9     & 38.0    &$-$1.0    & 30.8    &$-$14.29   & 6.73         & \multicolumn{2}{c}{--}   &  STD   \\
C/1978 H1       & 1.292    & 0.198   & 0.415    & 0.125   & \multicolumn{2}{c}{--}   & \multicolumn{2}{c}{--}   &  STD   \\
C/1986 P1-A     & 1.750    & 0.0600  & 0.0034   & 0.0271  & \multicolumn{2}{c}{--}   & \multicolumn{2}{c}{--}   &  STD   \\
C/1989 Q1       & 3.396    & 0.317   & 0.506    & 0.149   & \multicolumn{2}{c}{--}   & \multicolumn{2}{c}{--}   &  STD   \\
C/1989 X1       & 0.2675   & 0.0729  & 0.0327   & 0.0221  & \multicolumn{2}{c}{--}   & \multicolumn{2}{c}{--}   &  STD   \\
C/1990 K1       & 3.694    & 0.163   &$-$0.0736 & 0.0769  &   0.3001  & 0.0122       &   5.3  & 1.7             &  STD   \\
C/1991 F2       & 0.656    & 0.230   &$-$0.340  & 0.134   &   0.0724  & 0.0379       & \multicolumn{2}{c}{--}   &  STD   \\
C/1993 A1       & 14.458   & 0.215   & 0.135    & 0.279   &$-$0.0890  & 0.0627       &$-$34.2 & 3.2             &  STD   \\
C/1993 Q1       & 3.829    & 0.336   & 0.864    & 0.128   &   0.455   & 0.186        & \multicolumn{2}{c}{--}   &  STD   \\
C/1996 E1       & 6.688    & 0.311   & 0.6882   & 0.0935  & \multicolumn{2}{c}{--}   & \multicolumn{2}{c}{--}   &  STD   \\
C/1997 J2       & 292.74   & 6.92    &$-$50.39  & 6.35    & \multicolumn{2}{c}{--}   & \multicolumn{2}{c}{--}   &  STD   \\
C/1999 Y1       & 296.45   & 7.64    & 48.9     & 10.4    &   5.31    & 3.40         & \multicolumn{2}{c}{--}   &  STD   \\
C/2001 Q4       & 1.6506   & 0.0140  & 0.06240  & 0.00610 &   0.0141  & 0.0511       & \multicolumn{2}{c}{--}   &  STD   \\
                & 1.2132   & 0.0267  & 0.03323  & 0.00721 &   0.17907 & 0.00920      & \multicolumn{2}{c}{--}   &  DIST  \\
C/2002 E2       & 2.780    & 0.232   & 0.879    & 0.194   & \multicolumn{2}{c}{--}   & \multicolumn{2}{c}{--}   &  STD   \\
C/2002 T7       & 0.3822   & 0.0158  & 0.30233  & 0.00416 &$-$0.15256 & 0.00433      & \multicolumn{2}{c}{--}   &  STD   \\
                & 5.582    & 0.583   & 1.632    & 0.498   &   0.679   & 0.154        & \multicolumn{2}{c}{--}   &  PRE   \\
                & 1.414    & 0.106   & 0.4371   & 0.0781  &$-$0.4016  & 0.0292       & \multicolumn{2}{c}{--}   &  POST  \\
C/2003 K4       & 0.8522   & 0.0171  &$-$0.43620& 0.00607 &$-$0.06644 & 0.00432      & \multicolumn{2}{c}{--}   &  STD   \\
                & 0.3299   & 0.0407  & 0.1058   & 0.0256  &$-$0.1252  & 0.0148       & \multicolumn{2}{c}{--}   &  DIST  \\
C/2004 B1       & 0.9131   & 0.0334  &$-$0.3503 & 0.0177  &$-$0.23725 & 0.00683      & \multicolumn{2}{c}{--}   &  STD   \\
\multicolumn{10}{c}{{\bf S a m p l e ~~~A2}} \\
C/2006 K3       & 15.69    & 1.67    &  2.25    & 2.45    &$-$0.209   & 0.576        & \multicolumn{2}{c}{--}   &  STD   \\
C/2006 OF$_{2}$ & 2.384    & 0.168   &$-$1.370  & 0.131   &$-$0.0059  & 0.0347       & \multicolumn{2}{c}{--}   &  STD   \\
C/2006 P1       & 0.1329   & 0.0335  & 0.03138  & 0.00397 & \multicolumn{2}{c}{--}   & \multicolumn{2}{c}{--}   &  STD   \\
C/2006 Q1       & 33.504   & 0.700   & 1.916    & 0.550   & 10.604   & 0.189         & \multicolumn{2}{c}{--}   &  STD   \\
C/2006 VZ$_{13}$& 4.576    & 0.115   &$-$3.041  & 0.135   & 1.2204   & 0.0741        & \multicolumn{2}{c}{--}   &  STD   \\
                & 1.874    & 0.804   &$-$0.866  & 0.483   & 0.528    & 0.404         & \multicolumn{2}{c}{--}   &  PRE   \\
C/2007 N3       & 0.08678  & 0.00814 &$-$0.02141& 0.00696 &$-$0.13334 & 0.00190      &  11.3  & 1.9             &  STD   \\
C/2007 Q3       & 0.156    & 0.180   & 2.675    & 0.103   & 1.657     & 0.037        & \multicolumn{2}{c}{--}   &  STD   \\
C/2007 W1       & 3.97480  & 0.00841 &$-$0.4266 & 0.0138  &$-$0.06387 & 0.00276      & 24.7   & 4.3             &  STD   \\
                & 1.002    & 0.139   &$-$0.7253 & 0.0321  &$-$0.4916  & 0.0703       & \multicolumn{2}{c}{--}   &  PRE   \\
                & 5.864    & 0.272   &$-$0.783  & 0.172   &   0.136   & 0.250        & \multicolumn{2}{c}{--}   &  POST  \\
C/2007 W3       & 4.968    & 0.572   & 2.248    & 0.581   &$-$1.084   & 0.316        & \multicolumn{2}{c}{--}   &  STD   \\
C/2008 A1       & 5.5964   & 0.0570  & 0.7136   & 0.0384  & 0.16800   & 0.00815      & 5.76   & 0.61            &  STD   \\
                & 4.608    & 0.136   & 1.894    & 0.233   & 1.844     & 0.203        & \multicolumn{2}{c}{--}   &  PRE   \\
                &10.094    & 0.282   & 6.142    & 0.291   &$-$4.431   & 0.306        & \multicolumn{2}{c}{--}   &  POST  \\
C/2009 R1       & 5.798    & 0.490   &$-$1.418  & 0.359   & 0.776     & 0.207        & \multicolumn{2}{c}{--}   &  STD   \\
\multicolumn{10}{c}{{\bf S a m p l e ~~~B}} \\ 
C/1980 E1       &  1095    &   181   &  535.89  &  93.1   & \multicolumn{2}{c}{--}   & \multicolumn{2}{c}{--}  &  STD    \\ 
C/1983 O1       &  2683    &   942   &  158.10  & 677     & \multicolumn{2}{c}{--}   & \multicolumn{2}{c}{--}  &  STD    \\
C/1984 W2       & 36844    & 23157   & \multicolumn{2}{c}{--} & \multicolumn{2}{c}{--}& \multicolumn{2}{c}{--} &  STD \\   
C/1997 BA$_6$   &  3341    &   118   &   24.3   &  54.1   &$-$29.8    & 11.7         & \multicolumn{2}{c}{--}  &  STD    \\ 
C/1999 H3       &  4112    &   193   & 3007     & 228     &$-$509.0   & 72.0         & \multicolumn{2}{c}{--}  &  STD    \\ 
C/2000 CT$_{54}$&   778.0  &    53.6 &   51.5   &  25.9   & \multicolumn{2}{c}{--}   & \multicolumn{2}{c}{--}  &  STD    \\ 
C/2000 SV$_{74}$&  6121.6  &    80.2 &  717.9   &  60.7   &$-$513.6   & 21.7         & \multicolumn{2}{c}{--}  &  STD    \\ 
C/2002 R3       & 17850    &  2640   & 5810     &3510     & \multicolumn{2}{c}{--}   & \multicolumn{2}{c}{--}  &  STD    \\ 
C/2005 B1       &    74.7  &    12.6 &$-$77.7   &   9.17  &$-$63.94  & 4.66          & \multicolumn{2}{c}{--}  &  STD    \\ 
C/2005 EL$_{173}$& 6602    &   773   &$-$7175   &  496    & \multicolumn{2}{c}{--}   & \multicolumn{2}{c}{--}  &  STD    \\ 
C/2005 K1       &  2515    &   741   &   184    &  762    & \multicolumn{2}{c}{--}   & \multicolumn{2}{c}{--}  &  STD    \\ 
C/2006 S2       &   772    &   299   &$-$167    &  199    & \multicolumn{2}{c}{--}   & \multicolumn{2}{c}{--}  &  STD    \\ 
\hline
\end{tabular}}
\end{table*}

\vfill\newpage

\section{Part\,III -- Original barycentric orbital elements}

%%%%%%%%%%%%%%%%%%%%%%%%%%%%%%%%%%%%%%%%%%%%%%%%%%%%%%%%%%%%%%%%%%%%%%%%%%%%%%
%	comets from Paper 1 & 2 with q < 3.1 au;  27 objects, original orbits
%%%%%%%%%%%%%%%%%%%%%%%%%%%%%%%%%%%%%%%%%%%%%%%%%%%%%%%%%%%%%%%%%%%%%%%%%%%%%%

%\begin{longtab}
\setlength\LTcapwidth{1.00\textwidth} % default: 4in (rather less than \textwidth...)
%\setlength\LTleft{0pt}               % default: \parindent
%\setlength\LTright{0pt}              % default: \fill
%{\footnotesize {
{\setlength{\tabcolsep}{2.5pt}{
\begin{longtable}{ccrrrrrrr}
\caption{\label{tab:orbit_original_27} Orbital elements of original barycentric orbits, i.e. before entering the planetary zone
 for 28 near-parabolic comets described as sample~A1 in Table~\ref{tab:orbital_quality108_list}.
The successive cols. signify 
$[1]$ -- comet designation, $[2]$ -- epoch, i.e. osculation date,
$[3]$ -- perihelion time [TT], $[4]$ -- perihelion distance, $[5]$
-- eccentricity, $[6]$ -- argument of perihelion (in degrees),
equinox 2000.0, $[7]$ -- longitude of the ascending node (in
degrees), equinox 2000.0, $[8]$ -- inclination (in degrees), equinox
2000.0,    $[9]$ -- inverse original semi-major axis in units of $10^{-6}$\,au$^{-1}$. }\\
\hline\hline
%\toprule
Comet      & Epoch         & T                   & q            &  e         & $\omega$      & $\Omega$    &  i         & $1/a_{\rm ori}$ \\
           & $[$yyyymmdd$]$&$[$yyyymmdd.dddddd$]$& $[$au$]$     &            &  $[$\degr$]$  & $[$\degr$]$ & $[$\degr$]$& $[10^{-6}$au$^{-1}]$   \\
$[1]$      & $[2]$       & $[3]$                 & $[4]$        & $[5]$      & $[6]$         & $[7]$       & $[8]$      & $[9]$          \\
%\midrule
\hline
\endfirsthead
\caption{continued.}\\
\hline\hline
%
%\endfirsthead
%
%\multicolumn{9}{c}%
%{{\bfseries \tablename\ \thetable{} -- continued from the previous page}}\\
%\toprule
Comet      & Epoch         & T                   & q            &  e         & $\omega$      & $\Omega$    &  i         & $1/a_{\rm ori}$ \\
           & $[$yyyymmdd$]$&$[$yyyymmdd.dddddd$]$& $[$au$]$     &            &  $[$\degr$]$  & $[$\degr$]$ & $[$\degr$]$& $[10^{-6}$au$^{-1}]$   \\
$[1]$      & $[2]$       & $[3]$                 & $[4]$        & $[5]$      & $[6]$         & $[7]$       & $[8]$      & $[9]$          \\
%\midrule
\hline
\endhead
\hline 
\endfoot
%
%\bottomrule
%\endlastfoot
%
\\   
\multicolumn{9}{c}{{\bf Original barycentric orbits based on the entire data sets}} \\
%\\
C/1885 X1 & 15880117 & 18860405.846922 &      0.63999379 &      0.99996093 &     126.676913 &      37.853210 &      82.756505 &       61.04 \\
          &          &   $\pm$0.000595 & $\pm$0.00000574 & $\pm$0.00001099 &  $\pm$0.000644 &  $\pm$0.000093 &  $\pm$0.000459 & $\pm$ 17.17 \\
C/1892 Q1 & 15940415 & 18921228.151231 &      0.97466329 &      0.99994340 &     252.541362 &     266.215494 &      24.770458 &       58.07 \\
          &          &   $\pm$0.001357 & $\pm$0.00001784 & $\pm$0.00004960 &  $\pm$0.001495 &  $\pm$0.000171 &  $\pm$0.000148 & $\pm$ 50.89 \\
%C/1913 Y1 & 16151111 & 19141025.817363 &      1.10067114 &      0.99994214 &      97.541763 &      60.475772 &      67.986634 &       52.57 \\
%          &          &   $\pm$0.000089 & $\pm$0.00000460 & $\pm$0.00000465 &  $\pm$0.000303 &  $\pm$0.000024 &  $\pm$0.000059 & $\pm$  4.23 \\
%C/1940 R2 & 16430617 & 19410115.973647 &      0.36427385 &      0.99998121 &     199.547949 &     296.892704 &      50.184132 &       51.59 \\
%          &          &   $\pm$0.000186 & $\pm$0.00000275 & $\pm$0.00000624 &  $\pm$0.000223 &  $\pm$0.000181 &  $\pm$0.000151 & $\pm$ 17.12 \\
%C/1946 U1 & 16460313 & 19470206.863103 &      2.41082406 &      0.99993893 &     348.609219 &      35.567057 &     108.115230 &       25.33 \\
%          &          &   $\pm$0.002549 & $\pm$0.00001214 & $\pm$0.00001253 &  $\pm$0.000997 &  $\pm$0.000053 &  $\pm$0.000047 & $\pm$  5.20 \\
C/1952 W1 & 16550415 & 19530124.009962 &      0.77523403 &      1.00000006 &     253.927686 &     343.653652 &      97.162914 &       -0.07 \\
          &          &   $\pm$0.001204 & $\pm$0.00003670 & $\pm$0.00006651 &  $\pm$0.003436 &  $\pm$0.000900 &  $\pm$0.000824 & $\pm$ 85.79 \\
\bottomrule
\end{longtable}
%}}
}}
%\end{longtab}

%\newpage

%%%%%%%%%%%%%%%%%%%%%%%%%%%%%%%%%%%%%%%%%%%%%%%%%%%%%%%%%%%%%%%%%%%%%%%%%%%%%%
%	comets from Paper 4 with q < 3.1 au;  22 objects, original orbits
%%%%%%%%%%%%%%%%%%%%%%%%%%%%%%%%%%%%%%%%%%%%%%%%%%%%%%%%%%%%%%%%%%%%%%%%%%%%%%

%\begin{longtab}
\setlength\LTcapwidth{1.00\textwidth} % default: 4in (rather less than \textwidth...)
%\setlength\LTleft{0pt}               % default: \parindent
%\setlength\LTright{0pt}              % default: \fill
%{\footnotesize {
{\setlength{\tabcolsep}{2.5pt}{
\begin{longtable}{ccrrrrrrr}
\caption{\label{tab:orbit_original_22} Orbital elements of original barycentric orbits, i.e. before entering the planetary zone
 for 22 near-parabolic comets described as sample~A2 in Table~\ref{tab:orbital_quality108_list}.
The successive cols. signify
$[1]$ -- comet designation, $[2]$ -- epoch, i.e. osculation date,
$[3]$ -- perihelion time [TT], $[4]$ -- perihelion distance, $[5]$
-- eccentricity, $[6]$ -- argument of perihelion (in degrees),
equinox 2000.0, $[7]$ -- longitude of the ascending node (in
degrees), equinox 2000.0, $[8]$ -- inclination (in degrees), equinox
2000.0,    $[9]$ -- inverse original semi-major axis in units of $10^{-6}$\,au$^{-1}$. }\\
\hline\hline
%\toprule
Comet      & Epoch         & T                   & q            &  e         & $\omega$      & $\Omega$    &  i         & $1/a_{\rm ori}$ \\
           & $[$yyyymmdd$]$&$[$yyyymmdd.dddddd$]$& $[$au$]$     &            &  $[$\degr$]$  & $[$\degr$]$ & $[$\degr$]$& $[10^{-6}$au$^{-1}]$   \\
$[1]$      & $[2]$       & $[3]$                 & $[4]$        & $[5]$      & $[6]$         & $[7]$       & $[8]$      & $[9]$          \\
%\midrule
\hline
\endfirsthead
\caption{continued.}\\
\hline\hline
%
%\multicolumn{9}{c}%
%{{\bfseries \tablename\ \thetable{} -- continued from the previous page}}\\
%\toprule
Comet      & Epoch         & T                   & q            &  e         & $\omega$      & $\Omega$    &  i         & $1/a_{\rm ori}$ \\
           & $[$yyyymmdd$]$&$[$yyyymmdd.dddddd$]$& $[$au$]$     &            &  $[$\degr$]$  & $[$\degr$]$ & $[$\degr$]$& $[10^{-6}$au$^{-1}]$   \\
$[1]$      & $[2]$       & $[3]$                 & $[4]$        & $[5]$      & $[6]$         & $[7]$       & $[8]$      & $[9]$          \\
%\midrule
\hline
\endhead
\hline 
\endfoot
%
%\bottomrule
%\endlastfoot
%%
%\\
\multicolumn{9}{c}{{\bf Original barycentric orbits based on the entire data sets}} \\
C/2006 HW$_{51}$&17051119 & 20060929.574079 & 2.25933969 &      0.99989312 &       0.105489 &     228.163370 &      45.856715 &       47.31 \\
          &          &   $\pm$0.000189 & $\pm$0.00000070 & $\pm$0.00000762 &  $\pm$0.000043 &  $\pm$0.000038 &  $\pm$0.000023 & $\pm$  3.37 \\
C/2006 K3 & 16110913 & 20070313.052038 &      2.50325995 &      0.99984724 &     328.074663 &      49.427949 &      92.612447 &       61.02 \\
          &          &   $\pm$0.000278 & $\pm$0.00000304 & $\pm$0.00001158 &  $\pm$0.000153 &  $\pm$0.000043 &  $\pm$0.000018 & $\pm$  4.63 \\
C/2006 L2 & 17061114 & 20061119.409644 &      1.98559713 &      0.99997441 &      48.229422 &     239.260861 &     101.044452 &       12.89 \\
          &          &   $\pm$0.000117 & $\pm$0.00000069 & $\pm$0.00000283 &  $\pm$0.000042 &  $\pm$0.000012 &  $\pm$0.000023 & $\pm$  1.43 \\
C/2006 OF$_{2}$&17070930 & 20080914.944190 &  2.42852246 &      0.99994849 &      95.624276 &     318.584757 &      30.144295 &       21.21 \\
\bottomrule
\end{longtable}
%}}
}}
%\end{longtab}

%\newpage

%%%%%%%%%%%%%%%%%%%%%%%%%%%%%%%%%%%%%%%%%%%%%%%%%%%%%%%%%%%%%%%%%%%%%%%%%%%%%%
%	comets from Papers 1 & 2 with q > 3.1 au;  59 objects, original orbits
%%%%%%%%%%%%%%%%%%%%%%%%%%%%%%%%%%%%%%%%%%%%%%%%%%%%%%%%%%%%%%%%%%%%%%%%%%%%%%

%\begin{longtab}
\setlength\LTcapwidth{1.00\textwidth} % default: 4in (rather less than \textwidth...)
%\setlength\LTleft{0pt}               % default: \parindent
%\setlength\LTright{0pt}              % default: \fill
%{\footnotesize {
{\setlength{\tabcolsep}{2.5pt}{
\begin{longtable}{ccrrrrrrr}
\caption{\label{tab:orbit_original_59} Orbital elements of original barycentric orbits, i.e. before entering the planetary zone
 for 69 near-parabolic comets  described as sample~B in Table~\ref{tab:orbital_quality108_list}. All original orbits in this table are based on the osculating orbits determined from the entire data sets.
The successive cols. signify 
$[1]$ -- comet designation, $[2]$ -- epoch, i.e. osculation date,
$[3]$ -- perihelion time [TT], $[4]$ -- perihelion distance, $[5]$
-- eccentricity, $[6]$ -- argument of perihelion (in degrees),
equinox 2000.0, $[7]$ -- longitude of the ascending node (in
degrees), equinox 2000.0, $[8]$ -- inclination (in degrees), equinox
2000.0,    $[9]$ -- inverse original semi-major axis in units of $10^{-6}$\,au$^{-1}$. }\\
\hline\hline
%\toprule
Comet      & Epoch         & T                   & q            &  e         & $\omega$      & $\Omega$    &  i         & $1/a_{\rm ori}$ \\
           & $[$yyyymmdd$]$&$[$yyyymmdd.dddddd$]$& $[$au$]$     &            &  $[$\degr$]$  & $[$\degr$]$ & $[$\degr$]$& $[10^{-6}$au$^{-1}]$   \\
$[1]$      & $[2]$       & $[3]$                 & $[4]$        & $[5]$      & $[6]$         & $[7]$       & $[8]$      & $[9]$          \\
\hline
\endfirsthead
\caption{continued.}\\
\hline\hline
%
%\multicolumn{9}{c}%
%{{\bfseries \tablename\ \thetable{} -- continued from the previous page}}\\
%\toprule
Comet      & Epoch         & T                   & q            &  e         & $\omega$      & $\Omega$    &  i         & $1/a_{\rm ori}$ \\
           & $[$yyyymmdd$]$&$[$yyyymmdd.dddddd$]$& $[$au$]$     &            &  $[$\degr$]$  & $[$\degr$]$ & $[$\degr$]$& $[10^{-6}$au$^{-1}]$   \\
$[1]$      & $[2]$       & $[3]$                 & $[4]$        & $[5]$      & $[6]$         & $[7]$       & $[8]$      & $[9]$          \\
%\midrule
\hline
\endhead
\hline 
\endfoot
%
%\bottomrule
%\endlastfoot
%%
\\
C/1972 L1 & 16680605 & 19721115.398710 &      4.26458890 &      0.99978206 &      56.882790 &     225.480163 &      79.487939 &       51.11 \\
          &          &   $\pm$0.003935 & $\pm$0.00001844 & $\pm$0.00002648 &  $\pm$0.000579 &  $\pm$0.000117 &  $\pm$0.000073 & $\pm$  6.21 \\
C/1973 W1 & 16690819 & 19730810.055023 &      3.84026288 &      0.99972033 &     221.390019 &     244.629964 &     108.049930 &       72.83 \\
          &          &   $\pm$0.006665 & $\pm$0.00008612 & $\pm$0.00004727 &  $\pm$0.002037 &  $\pm$0.000278 &  $\pm$0.000094 & $\pm$ 12.31 \\
C/1974 V1 & 16670721 & 19740808.164070 &      6.01579264 &      0.99989464 &     151.829672 &     226.121112 &      60.878468 &       17.51 \\
          &          &   $\pm$0.019246 & $\pm$0.00009240 & $\pm$0.00007279 &  $\pm$0.001856 &  $\pm$0.000068 &  $\pm$0.000198 & $\pm$ 12.10 \\
C/1976 D2 & 16660107 & 19750115.346261 &      6.88171236 &      0.99960808 &     193.436940 &      22.762585 &     111.987797 &       56.95 \\
\bottomrule
\end{longtable}
%}}
}}
%\end{longtab}

\newpage

\section{Part\,IV -- Future barycentric orbital elements}

%%%%%%%%%%%%%%%%%%%%%%%%%%%%%%%%%%%%%%%%%%%%%%%%%%%%%%%%%%%%%%%%%%%%%%%%%%%%%%
%	comets from  with q < 3.1 au;  28 objects, future orbits
%%%%%%%%%%%%%%%%%%%%%%%%%%%%%%%%%%%%%%%%%%%%%%%%%%%%%%%%%%%%%%%%%%%%%%%%%%%%%%

%\begin{longtab}
\setlength\LTcapwidth{1.00\textwidth} % default: 4in (rather less than \textwidth...)
%\setlength\LTleft{0pt}               % default: \parindent
%\setlength\LTright{0pt}              % default: \fill
%{\footnotesize {
{\setlength{\tabcolsep}{2.5pt}{
\begin{longtable}{ccrrrrrrr}
\caption{\label{tab:orbit_future_27} Orbital elements of future barycentric orbits, i.e. after leaving the planetary zone
 for 28 near-parabolic comets described as sample~A1 in Table~\ref{tab:orbital_quality108_list}.
The successive cols. signify
$[1]$ -- comet designation, $[2]$ -- epoch, i.e. osculation date,
$[3]$ -- perihelion time [TT], $[4]$ -- perihelion distance, $[5]$
-- eccentricity, $[6]$ -- argument of perihelion (in degrees),
equinox 2000.0, $[7]$ -- longitude of the ascending node (in
degrees), equinox 2000.0, $[8]$ -- inclination (in degrees), equinox
2000.0,    $[9]$ -- inverse future semi-major axis in units of $10^{-6}$\,au$^{-1}$. }\\
\hline\hline
%\toprule
Comet      & Epoch         & T                   & q            &  e         & $\omega$      & $\Omega$    &  i         & $1/a_{\rm fut}$ \\
           & $[$yyyymmdd$]$&$[$yyyymmdd.dddddd$]$& $[$au$]$     &            &  $[$\degr$]$  & $[$\degr$]$ & $[$\degr$]$& $[10^{-6}$au$^{-1}]$   \\
$[1]$      & $[2]$       & $[3]$                 & $[4]$        & $[5]$      & $[6]$         & $[7]$       & $[8]$      & $[9]$          \\
%\midrule
\hline
\endfirsthead
\caption{continued.}\\
\hline\hline
%
%\multicolumn{9}{c}%
%{{\bfseries \tablename\ \thetable{} -- continued from the previous page}}\\
%\toprule
Comet      & Epoch         & T                   & q            &  e         & $\omega$      & $\Omega$    &  i         & $1/a_{\rm fut}$ \\
           & $[$yyyymmdd$]$&$[$yyyymmdd.dddddd$]$& $[$au$]$     &            &  $[$\degr$]$  & $[$\degr$]$ & $[$\degr$]$& $[10^{-6}$au$^{-1}]$   \\
$[1]$      & $[2]$       & $[3]$                 & $[4]$        & $[5]$      & $[6]$         & $[7]$       & $[8]$      & $[9]$          \\
\hline
\endhead
\hline 
\endfoot
%
%\bottomrule
%\endlastfoot
%%
\\   
\multicolumn{9}{c}{{\bf Future barycentric orbits based on the entire data sets}} \\
C/1885 X1 & 21820810 & 18860407.057135 &      0.64058887 &      1.00006625 &     126.518354 &      38.136646 &      82.607618 &     -103.42 \\
          &          &   $\pm$0.000133 & $\pm$0.00000783 & $\pm$0.00002647 &  $\pm$0.001336 &  $\pm$0.000094 &  $\pm$0.000459 & $\pm$ 41.32 \\
C/1892 Q1 & 21851013 & 18921228.159564 &      0.97762751 &      1.00047597 &     252.717846 &     265.987063 &      24.766253 &     -486.86 \\
          &          &   $\pm$0.001086 & $\pm$0.00000825 & $\pm$0.00002276 &  $\pm$0.000772 &  $\pm$0.000186 &  $\pm$0.000149 & $\pm$ 23.28 \\
%C/1913 Y1 & 22140224 & 19141026.924707 &      1.10515434 &      0.99990403 &      97.458874 &      60.292937 &      67.982497 &       86.84 \\
%          &          &   $\pm$0.000072 & $\pm$0.00000460 & $\pm$0.00000467 &  $\pm$0.000300 &  $\pm$0.000023 &  $\pm$0.000059 & $\pm$  4.23 \\
%C/1940 R2 & 22221129 & 19410115.373155 &      0.37020004 &      1.00055310 &     199.639408 &     296.591043 &      49.669507 &    -1494.06 \\
%          &          &   $\pm$0.000183 & $\pm$0.00000311 & $\pm$0.00001323 &  $\pm$0.000646 &  $\pm$0.000184 &  $\pm$0.000154 & $\pm$ 35.72 \\
%C/1946 U1 & 22480426 & 19470207.498214 &      2.41249282 &      0.99987424 &     348.623514 &      35.534123 &     108.158751 &       52.13 \\
%          &          &   $\pm$0.001146 & $\pm$0.00001215 & $\pm$0.00001253 &  $\pm$0.000477 &  $\pm$0.000052 &  $\pm$0.000046 & $\pm$  5.19 \\
C/1952 W1 & 22500526 & 19530125.074189 &      0.77661881 &      1.00003215 &     253.740607 &     343.512621 &      97.122335 &      -41.39 \\
          &          &   $\pm$0.002219 & $\pm$0.00004215 & $\pm$0.00008305 &  $\pm$0.004184 &  $\pm$0.000894 &  $\pm$0.000825 & $\pm$106.94 \\
\bottomrule
\end{longtable}
%}}
}}
%\end{longtab}

%\newpage

%%%%%%%%%%%%%%%%%%%%%%%%%%%%%%%%%%%%%%%%%%%%%%%%%%%%%%%%%%%%%%%%%%%%%%%%%%%%%%
%	comets from Paper 4 with q < 3.1 au;  22 objects future orbits
%%%%%%%%%%%%%%%%%%%%%%%%%%%%%%%%%%%%%%%%%%%%%%%%%%%%%%%%%%%%%%%%%%%%%%%%%%%%%%

%\begin{longtab}
\setlength\LTcapwidth{1.00\textwidth} % default: 4in (rather less than \textwidth...)
%\setlength\LTleft{0pt}               % default: \parindent
%\setlength\LTright{0pt}              % default: \fill
%{\footnotesize {
{\setlength{\tabcolsep}{2.5pt}{
\begin{longtable}{ccrrrrrrr}
\caption{\label{tab:orbit_future_22} Orbital elements of future barycentric orbits, i.e. after leaving the planetary zone
 for 21 near-parabolic comets described as sample~A2 in Table~\ref{tab:orbital_quality108_list} (comet C/2010~X1 disintegrated shortly after perihelion passage).
The successive cols. signify
$[1]$ -- comet designation, $[2]$ -- epoch, i.e. osculation date,
$[3]$ -- perihelion time [TT], $[4]$ -- perihelion distance, $[5]$
-- eccentricity, $[6]$ -- argument of perihelion (in degrees),
equinox 2000.0, $[7]$ -- longitude of the ascending node (in
degrees), equinox 2000.0, $[8]$ -- inclination (in degrees), equinox
2000.0,    $[9]$ -- inverse future semi-major axis in units of $10^{-6}$\,au$^{-1}$. }\\
\hline\hline
%\toprule
Comet      & Epoch         & T                   & q            &  e         & $\omega$      & $\Omega$    &  i         & $1/a_{\rm fut}$ \\
           & $[$yyyymmdd$]$&$[$yyyymmdd.dddddd$]$& $[$au$]$     &            &  $[$\degr$]$  & $[$\degr$]$ & $[$\degr$]$& $[10^{-6}$au$^{-1}]$   \\
$[1]$      & $[2]$       & $[3]$                 & $[4]$        & $[5]$      & $[6]$         & $[7]$       & $[8]$      & $[9]$          \\
\hline
\endfirsthead
\caption{continued.}\\
\hline\hline
%\midrule
%\endfirsthead
%
%\multicolumn{9}{c}%
%{{\bfseries \tablename\ \thetable{} -- continued from the previous page}}\\
%\toprule
Comet      & Epoch         & T                   & q            &  e         & $\omega$      & $\Omega$    &  i         & $1/a_{\rm fut}$ \\
           & $[$yyyymmdd$]$&$[$yyyymmdd.dddddd$]$& $[$au$]$     &            &  $[$\degr$]$  & $[$\degr$]$ & $[$\degr$]$& $[10^{-6}$au$^{-1}]$   \\
$[1]$      & $[2]$       & $[3]$                 & $[4]$        & $[5]$      & $[6]$         & $[7]$       & $[8]$      & $[9]$          \\
\hline
\endhead
\hline 
\endfoot
%
%\bottomrule
%\endlastfoot
%%
%\\
\multicolumn{9}{c}{{\bf Future barycentric orbits based on the entire data sets}} \\
C/2006 HW$_{51}$&23080212 & 20060929.188535 & 2.25852845 &      0.99979646 &     359.781513 &     228.103038 &      45.851873 &       90.12 \\
          &          &   $\pm$0.000129 & $\pm$0.00000068 & $\pm$0.00000762 &  $\pm$0.000044 &  $\pm$0.000037 &  $\pm$0.000023 & $\pm$  3.37 \\
C/2006 K3 & 23060622 & 20070313.561543 &      2.50568081 &      1.00032895 &     328.094463 &      49.406950 &      92.578493 &     -131.28 \\
          &          &   $\pm$0.000802 & $\pm$0.00000437 & $\pm$0.00001170 &  $\pm$0.000239 &  $\pm$0.000028 &  $\pm$0.000053 & $\pm$  4.67 \\
C/2006 L2 & 23050915 & 20061119.940263 &      1.99185210 &      1.00018974 &      47.969720 &     239.288007 &     101.012257 &      -95.26 \\
          &          &   $\pm$0.000125 & $\pm$0.00000070 & $\pm$0.00000284 &  $\pm$0.000042 &  $\pm$0.000012 &  $\pm$0.000023 & $\pm$  1.43 \\
C/2006 OF$_{2}$& 23020315 & 20080915.318008 & 2.43768516 &      1.00160599 &      95.754006 &     318.447571 &      30.112628 &     -658.82 \\
\bottomrule
\end{longtable}
%}}
}}
%\end{longtab}

%\newpage

%%%%%%%%%%%%%%%%%%%%%%%%%%%%%%%%%%%%%%%%%%%%%%%%%%%%%%%%%%%%%%%%%%%%%%%%%%%%%%
%	comets with q > 3.1 au;  69 objects, future orbits
%%%%%%%%%%%%%%%%%%%%%%%%%%%%%%%%%%%%%%%%%%%%%%%%%%%%%%%%%%%%%%%%%%%%%%%%%%%%%%

%\begin{longtab}
\setlength\LTcapwidth{1.00\textwidth} % default: 4in (rather less than \textwidth...)
{\setlength{\tabcolsep}{2.5pt}{
\begin{longtable}{ccrrrrrrr}
\caption{\label{tab:orbit_future_59} Orbital elements of future barycentric orbits, i.e. after leaving the planetary zone
 for 69 near-parabolic comets  described as sample~B in Table~\ref{tab:orbital_quality108_list}. All future orbits in this table are based on the osculating orbits determined from the entire data sets.
The successive cols. signify 
$[1]$ -- comet designation, $[2]$ -- epoch, i.e. osculation date,
$[3]$ -- perihelion time [TT], $[4]$ -- perihelion distance, $[5]$
-- eccentricity, $[6]$ -- argument of perihelion (in degrees),
equinox 2000.0, $[7]$ -- longitude of the ascending node (in
degrees), equinox 2000.0, $[8]$ -- inclination (in degrees), equinox
2000.0,    $[9]$ -- inverse future semi-major axis in units of $10^{-6}$\,au$^{-1}$. }\\
\hline\hline
%\toprule
Comet      & Epoch         & T                   & q            &  e         & $\omega$      & $\Omega$    &  i         & $1/a_{\rm fut}$ \\
           & $[$yyyymmdd$]$&$[$yyyymmdd.dddddd$]$& $[$au$]$     &            &  $[$\degr$]$  & $[$\degr$]$ & $[$\degr$]$& $[10^{-6}$au$^{-1}]$   \\
$[1]$      & $[2]$       & $[3]$                 & $[4]$        & $[5]$      & $[6]$         & $[7]$       & $[8]$      & $[9]$          \\
\hline
\endfirsthead
\caption{continued.}\\
\hline\hline
%\midrule
%\endfirsthead
%
%\multicolumn{9}{c}%
%{{\bfseries \tablename\ \thetable{} -- continued from the previous page}}\\
%\toprule
Comet      & Epoch         & T                   & q            &  e         & $\omega$      & $\Omega$    &  i         & $1/a_{\rm fut}$ \\
           & $[$yyyymmdd$]$&$[$yyyymmdd.dddddd$]$& $[$au$]$     &            &  $[$\degr$]$  & $[$\degr$]$ & $[$\degr$]$& $[10^{-6}$au$^{-1}]$   \\
$[1]$      & $[2]$       & $[3]$                 & $[4]$        & $[5]$      & $[6]$         & $[7]$       & $[8]$      & $[9]$          \\
\hline
\endhead
\hline 
\endfoot
%
%\bottomrule
%\endlastfoot
%%
\\
C/1972 L1 & 22691122 & 19721113.990154 &      4.27440651 &      1.00265385 &      56.572391 &     225.472679 &      79.414662 &     -620.87 \\
          &          &   $\pm$0.003911 & $\pm$0.00001806 & $\pm$0.00002653 &  $\pm$0.000589 &  $\pm$0.000117 &  $\pm$0.000072 & $\pm$  6.20 \\
C/1973 W1 & 22750624 & 19730810.219208 &      3.84312446 &      1.00041071 &     221.296365 &     244.622483 &     108.018234 &     -106.87 \\
          &          &   $\pm$0.006610 & $\pm$0.00008654 & $\pm$0.00004732 &  $\pm$0.002035 &  $\pm$0.000278 &  $\pm$0.000096 & $\pm$ 12.31 \\
C/1974 V1 & 22880705 & 19740806.921353 &      6.01222855 &      0.99654636 &     151.677213 &     226.099858 &      60.909106 &      574.44 \\
          &          &   $\pm$0.019374 & $\pm$0.00009311 & $\pm$0.00007283 &  $\pm$0.001868 &  $\pm$0.000068 &  $\pm$0.000199 & $\pm$ 12.12 \\
C/1976 D2 & 22840108 & 19750115.363704 &      6.88351258 &      0.99962729 &     193.402971 &      22.777758 &     111.985520 &       54.15 \\
\bottomrule
\end{longtable}
}}
%\end{longtab}

}}


\begin{thebibliography}{}

\bibitem[\protect\citeauthoryear{Dybczy{\'n}ski \&
  Kr{\'o}likowska}{Dybczy{\'n}ski \& Kr{\'o}likowska}{2011}]{dyb-kroli:2011}
Dybczy{\'n}ski P.~A.,  Kr{\'o}likowska M.,  2011, MNRAS, 416, 51 (Paper 2)

\bibitem[\protect\citeauthoryear{IAU~Minor Planet Center Database}{IAU~Minor Planet Center Database}{2013}]{IAU_MPC_Web}
IAU~Minor~Planet~Center, 2013, MPC~Database~Search, URL
http://www.minorplanetcenter.net/db\_search/

\bibitem[\protect\citeauthoryear{IAU~Minor Planet Center Web Pages}{IAU~Minor Planet Center Web Pages}{2013}]{IAU_MPC_Web2}
IAU~Minor~Planet~Center, 2013, Uncertainty Parameter U and Orbit Quality Codes, URL
http://www.minorplanetcenter.net/iau/info/UValue.html 

\bibitem[\protect\citeauthoryear{JPL Small-Body Database Browser}{JPL Small-Body Database Browser}{2013}]{JPL_Browser}
JPL Small-Body Database Browser, 2013, JPL~Database~Search, URL
http://ssd.jpl.nasa.gov/sbdb.cgi 

\bibitem[\protect\citeauthoryear{Kr{\'o}likowska}{Kr{\'o}likowska}{2006}]{krolikowska:2006a}
Kr{\'o}likowska M.,  2006, Acta Astronomica, 56, 385

\bibitem[\protect\citeauthoryear{Kr{\'o}likowska \&
  Dybczy{\'n}ski}{Kr{\'o}likowska \& Dybczy{\'n}ski}{2010}]{kroli-dyb:2010}
Kr{\'o}likowska M.,  Dybczy{\'n}ski P.~A.,  2010, MNRAS, 404, 1886 (Paper 1)

\bibitem[\protect\citeauthoryear{{Kr{\'o}likowska}, {Dybczy{\'n}ski} \&
  {Sitarski}}{{Kr{\'o}likowska} et~al.}{2012}]{kroli-dyb:2012}
{Kr{\'o}likowska} M.,  {Dybczy{\'n}ski} P.~A.,    {Sitarski} G.,  2012, A\&A,
  544, A119 (Paper 3)

\bibitem[\protect\citeauthoryear{Kr{\'o}likowska \&
  Dybczy{\'n}ski}{Kr{\'o}likowska \& Dybczy{\'n}ski}{2013}]{kroli-dyb:2013}
Kr{\'o}likowska M.,  Dybczy{\'n}ski P.~A.,  2013, MNRAS, 435, 440 (Paper 4)

\bibitem[\protect\citeauthoryear{Marsden, Sekanina \& Everhart}{Marsden
  et~al.}{1978}]{mar-sek-eve:1978}
Marsden B.~G.,  Sekanina Z.,    Everhart E.,  1978, AJ, 83, 64

\bibitem[\protect\citeauthoryear{Marsden, Sekanina \& Yeomans}{Marsden
  et~al.}{1973}]{marsden-sek-ye:1973}
Marsden B.~G.,  Sekanina Z.,    Yeomans D.~K.,  1973, AJ, 78, 211

\bibitem[\protect\citeauthoryear{Marsden \& Williams Catalogue}{Marsden \& Williams Catalogue}{2008}]{MWC08:2008}
Marsden B.~G.,  Williams G.V., 2008, Catalogue of Cometary Orbits 17th Edition, Smithsonian Astrophysical
Observatory, Cambridge, Mass. 

\bibitem[\protect\citeauthoryear{Nakano}{Nakano}{2009a}]{NK1731A}
Nakano S., 2009, Nakano Note 1731a, URL
http://www.oaa.gr.jp/~oaacs/nk1731a.htm

\bibitem[\protect\citeauthoryear{Nakano}{Nakano}{2009b}]{NK1731B}
Nakano S., 2009, Nakano Note 1731b, URL
http://www.oaa.gr.jp/~oaacs/nk1731b.htm

\bibitem[\protect\citeauthoryear{Nakano}{Nakano}{2009c}]{NK1807}
Nakano S., 2009, Nakano Note 1807, URL
http://www.oaa.gr.jp/~oaacs/nk1807.htm

\bibitem[\protect\citeauthoryear{Sitarski}{Sitarski}{1989}]{sitarski:1989}
Sitarski G.,  1989, Acta Astronomica, 39, 345

\bibitem[\protect\citeauthoryear{Sitarski}{Sitarski}{1998}]{sitarski:1998}
Sitarski G.,  1998, Acta Astronomica, 48, 547

\bibitem[\protect\citeauthoryear{Sitarski}{Sitarski}{2002}]{sitarski:2002}
Sitarski G.,  2002, Acta Astronomica, 52, 471

\end{thebibliography}
\end{document}